\documentclass[runningheads,orivec]{llncs}
\usepackage[T1]{fontenc}
\usepackage{type1cm} 
\usepackage{graphicx} 
\usepackage{xspace} 
\usepackage{balance} 
\usepackage{booktabs} 
\usepackage{makecell}
\usepackage{multirow} 
\usepackage{longtable}
\usepackage[font={bf}, tableposition=top]{caption} 
\usepackage{subcaption} 
\usepackage{bold-extra} 
\usepackage[vlined,linesnumbered,ruled,noend]{algorithm2e} 
\usepackage{microtype} 
\usepackage{siunitx} 
\usepackage{xfrac} 
\usepackage{mathtools} 
\usepackage[hyphens]{url} 
\usepackage[bookmarks, pdftex, colorlinks=true, pagebackref=true, backref=page]{hyperref} 
\usepackage[square,numbers]{natbib} 
\usepackage{cleveref} 
\usepackage[hyperpageref]{backref} 
\usepackage{hyphenat} 
\usepackage[normalem]{ulem}

\usepackage{pgfplots}
\usepackage[all]{nowidow}
\usepackage[utf8]{inputenc}
\usepackage{tikz}
\usetikzlibrary{er,positioning,bayesnet,arrows.meta,calc}
\usepackage{multicol}
\usepackage[inline]{enumitem} 
\usepackage{MnSymbol}%
\usepackage{wasysym}%
\usepackage{diagbox}

\newcommand{\spara}[1]{\smallskip\noindent\textbf{#1}}

\renewcommand*\backref[1]{\ifx#1\relax \else (Cited on #1) \fi}

\usepackage{xcolor}
\hypersetup{
    bookmarks, pdftex,
    colorlinks=true,
    pagebackref=true, backref=page,
    linkcolor={red!50!black},
    filecolor={green!50!black},
    citecolor={green!50!black}, 
    urlcolor={blue!80!black},
}

\graphicspath{{figures}}

\definecolor{blue}{HTML}{1F77B4}
\definecolor{orange}{HTML}{FF7F0E}
\definecolor{green}{HTML}{2CA02C}
\definecolor{WildStrawberry}{HTML}{FF43A4}
\definecolor{BlueGreen}{RGB}{8,143,143}

\pgfplotsset{compat=1.14}

\makeatletter
\AtBeginDocument{%
  \def\doi#1{\url{https://doi.org/#1}}}
\makeatother

\newcommand{\R}[1]{\texttt{/r/#1}\xspace}

\begin{document}
\title{Moral Judgments in Online Discourse\\are not Biased by Gender}
%
%
\author{Lorenzo Betti\inst{1,2} \and
Paolo Bajardi\inst{3} \and
Gianmarco De Francisci Morales\inst{3}}
%

\institute{ISI Foundation, Turin, Italy \and Department of Network and Data Science, Central European University, Vienna, Austria \and
CENTAI, Turin, Italy\\
\email{lrn.betti@gmail.com} \\
\email{paolo.bajardi@centai.eu}\\
\email{gdfm@acm.org} }

\maketitle              
%
\begin{abstract} 

The interaction between social norms and gender roles prescribes gender-specific behaviors that influence moral judgments. 
Here, we study how moral judgments are biased by the gender of the protagonist of a story.
Using data from \R{AITA}, a Reddit community with 17 million members who share first-hand experiences seeking community judgment on their behavior, we employ machine learning techniques to match stories describing similar situations that differ only by the protagonist's gender. 
We find no direct causal effect of the protagonist's gender on the received moral judgments, except for stories about ``friendship and relationships'', where male protagonists receive more negative judgments.
Our findings complement existing correlational studies and suggest that gender roles may exert greater influence in specific social contexts.
These results have implications for understanding sociological constructs and highlight potential biases in data used to train large language models.

\end{abstract}

\section{Introduction}
\label{sec:intro}

Social norms are the informal rules that govern behavior in groups and societies, and represent collective beliefs about which behavior is appropriate in a given situation.
Extensively studied in the social sciences~\citep{durkheim1985regles, parsons1937structure, coleman1994foundations}, much is known about their formation, persistence, evolution, function, effects, and their link to social identity~\citep{tajfel1973roots}. 
These norms are intimately connected to social roles---behavioral archetypes that define specific expectations that individuals are presumed to fulfill when they embody certain roles within a social unit~\citep{fitzgerald2016social,sunstein1996social}.
In particular, gender roles, as a subclass of social roles, prescribe gender-specific behaviors~\cite{lindsey2015gender,pino2020hearty}.
For instance, hand-kissing as a form of greeting is a social norm intrinsically linked to gender roles, and serves as a vivid example of gender-specific behaviors.
The internalization of social and gender roles affects how individuals make moral judgments, thus influencing people's behavior via pressure to conform and sanctions such as isolation and ostracism~\citep{scott1971internalization}.
Investigating how social and gender roles influence moral judgments allows us to better understand human decision-making processes.

Despite efforts in studying the impact of judges' characteristics on the judgment, fewer works have studied how moral judgments are shaped by the characteristics of individuals involved in the moral dilemma.
For instance, it has been shown that old and female individuals are judged more mildly than young and male ones~\cite{reynolds2020man,chu2018moral}.
Such experiments employ vignettes, short stories based on fictional or hypothetical scenarios~\cite{mah2014using,christensen2012moral}, to expose participants to moral situations and ask for their judgment.
Although this methodology allows the experimenter to manipulate features of the moral dilemma to investigate how they modulate moral judgments, such as the demographics of the actors involved in the vignette, the stories may not reflect the complexity of real-world situations, and this may limit the generalizability of the findings.

The present study uses stories coming from a popular online social forum, Reddit, to address our main research question: \emph{``Does the declared gender of the protagonist of a story affect the moral judgment they receive''? }
In particular, the subreddit \R{AITA} (Am I The Asshole) is one of several communities dedicated to discussing conflicts that arise in everyday life and expressing moral judgments on these situations.
In this subreddit, users share stories describing a morally ambiguous situation that they experienced and how they behaved in it.
The community then decides whether the protagonist of the story was in the wrong, i.e., whether their behavior was deviant and violated a social norm.
This collection of moral judgments is invaluable to understanding social norms.
In particular, protagonists often self-disclose their demographic attributes (i.e., age and gender) and the community expresses their judgment by using a defined set of tags, thus enabling the collection of protagonists' demographics and the moral judgments they receive at scale.
This information can then be used to study the interplay between demographics and social norms across a wide spectrum of contexts.

Other studies have explored a similar setting on Reddit, albeit not in a causal setting~\citep{chandrasekharan2018internet,nguyen2022mapping,miller2020investigating}.
Specifically, multiple studies have observed the presence of a gender disparity in the moral judgments received on Reddit~\citep{Candia2022Social,botzer2022analysis}, where male and older protagonists are judged more negatively.
These studies are consistent with the observations that men are more easily perceived as perpetrators while women as victims~\citep{reynolds2020man,arnestad2020hetoo}, however, they are correlational.
While they propose several hypotheses about the causal mechanisms behind this imbalance, they do not provide any concrete evidence for them.
The present work addresses this gap: we design a causal observational study to understand the effect of the declared gender on the moral judgments received.
Compared to experimental studies, our design allows reaching a much larger sample that is still ecologically valid.
That is, rather than using synthetic or manipulated stories, we focus on the real-world narratives and judgments that originate from the community under study.

\begin{figure}[htbp]
\begin{center}
\begin{tikzpicture}

    \node (gender) [draw, rectangle, thick] {Gender};
    \node (age) [draw, rectangle, below=of gender] {Age};
    \node (situation) [draw, align=center, rectangle, right=2cm of $(gender)!0.5!(age)$] {Experiencing\\a situation};
    \node (submission) [draw, align=center, rectangle, thick, right=2cm of situation] {Posting on\\\texttt{r/AITA}};
    \node (judgment) [draw, rectangle, thick, right=2cm of submission] {Judgment};

    \draw [->] (gender) -- (situation);
    \draw [->] (age) -- (situation);

    \draw [->] (gender) to [bend left] (submission);
    \draw [->] (age) to [bend right] (submission);

    \draw [->,densely dotted] (gender) to [bend left] (judgment);
    \draw [->] (age) to [bend right] (judgment);

    \draw [->] (situation) -- (submission);
    \draw [->] (submission) -- (judgment);

\end{tikzpicture}
\vspace{\baselineskip}
\caption{
Causal graph encoding our hypotheses on the judgment mechanism in \R{AITA}. 
Given a submission, we assume that the type of situation experienced (``Experiencing a situation'') is influenced by the gender and age of the protagonist (``Gender'' and ``Age'').
Then, these three elements all affect the likelihood that the protagonist shares this situation on \R{AITA} (``Posting on \R{AITA}).
Finally, the moral judgment received (``Judgment'') is caused by the gender of the protagonist through two causal paths: directly, which reflects the first hypothesis, and mediated by the type of situation described in the submission, which refers to the second hypothesis.
The dashed edge from ``Gender'' to ``Judgment'' is the causal effect we aim to measure.
}
\label{fig:causal-dag}
\end{center}
\end{figure}
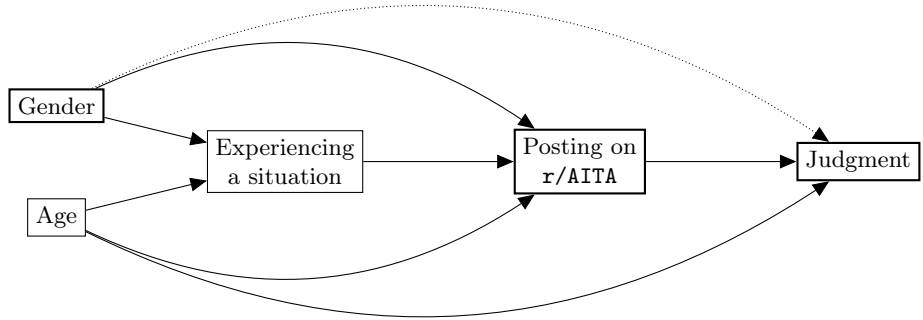

Two competing possible causes have been brought forward for the observed disparity in negative judgments received by men and women, which are represented in the causal graph depicted in \Cref{fig:causal-dag}.
The first hypothesis is that judges are biased by the gender of the protagonist of the story, e.g., by gender stereotypes or homophilic effects~\citep{chu2018moral,pino2020hearty}.
In this first scenario, the gender of the protagonist would have a direct effect on the judgment received (i.e., the dashed line from ``Gender'' to ``Judgment'' in \Cref{fig:causal-dag}).
If such a hypothesis were true, one would expect a bias according to which male protagonists receive more negative judgments, as reported in prior observational~\cite{Candia2022Social} and experimental studies~\cite{reynolds2020man}.
The second hypothesis is that men and women might be differently inclined to share morally ambiguous situations.
For instance, men are known to be more comfortable with risky situations~\citep{byrnes1999gender,barber2001boys,croson2009gender} while women tend to use online communities as support groups~\citep{klemm1999gender,tifferet2020gender,barbee1993effects}.
As a consequence, male protagonists may be more likely to receive negative judgments because of their greater propensity to share morally ambiguous situations.
In this second case, the reported situation acts as a mediator between the gender of the protagonist and the judgments received (i.e., the path from ``Gender'' to ``Judgment'' that passes through ``Posting on \R{AITA}'').
Therefore, if such a hypothesis were true, no effect of gender on moral judgments would be observed when measuring the effect of gender on moral judgments by averaging the situations out.
In other words, we wish to verify whether the situation is a partial or full mediator between gender and judgment.

To distinguish between these two competing hypotheses, our study design matches pairs of stories that describe similar situations but whose protagonists have different genders.
Thus, we can estimate the causal effect of the gender (treatment) on the judgment (outcome) by employing an approach based on propensity score matching~\citep{stuart2010matching}, which allows controlling for possible confounders.
We enforce matching between similar stories featuring male and female users by using advanced machine-learned language models and computing document embedding similarities.
We also control for other factors such as the age of the protagonist and the general topic of discussion.
This design has two main advantages.
First, by leveraging easily-collected, observational data we have access to a large sample of moral judgments.
Second, these stories are organically reported by the authors, and thus our results maintain ecological validity (which is harder in a laboratory study).
Finally, we hope that our study design can provide an example of how to tackle such important causal questions and inspire similar inquiries.

Our results indicate no significant direct causal effect between the gender of the protagonist and the judgment of the community.
When controlling for the situation described in the story, male protagonists are no more likely to receive a negative judgment than female ones.
However, when disaggregating the results by topic, we find a small but significant negative bias towards male protagonists in only one of them: ``friendship and relationships''.
Possibly, the difference in this specific topic is due to social norms that are more connected to gender roles than in other contexts~\citep{xi2023blame}.

The question tackled in this work has clear societal implications, as it sheds light on the interaction between fundamental sociological constructs.
The fact that this bias might be present within an online social platform such as Reddit has additional relevance in an AI-focused era, as large language models use Reddit extensively as training data.
If such causal bias were evident, then models trained on this data would be likely to reproduce and possibly amplify it further.

\section{Results}

We collect all submissions containing demographic information about their authors.
To assess the community judgment of these posts, we extract and count community-specific judgment tags in the comments, which indicate whether the protagonist was judged to deserve blame (see Methods for details).
This process results in a total of \num{33421} submissions annotated for judgment and demographics:
\num{21.612} authored by females ($65\%$), while \num{7269} judged to deserve blame ($22\%$).
Nearly half of the submissions are authored by users aged 19 to 26 for both male and female authors (see \Cref{fig:app_distrib_age_gender} in the Appendix).

\spara{Male authors are more likely to be judged deviant.} 
As described in \Cref{fig:causal-dag} (dotted line), we first investigate the association between gender and judgment.
Previous studies report a significant correlation whereby male and older authors are more likely to receive a negative judgment, although with different magnitudes~\cite{xi2023blame,giorgi2023author,Candia2022Social}.
Our crude estimate aligns with these studies: male authors are approximately twice as likely as female ones to receive a negative judgment from the community (OR~$= 2.21$, $95\%$ CI: $2.10$--$2.33$, Fisher's exact test $p\text{-value} < 0.001$).
This association is consistent across different topics, as determined by the Latent Dirichlet Allocation (LDA) topic model~\citep{blei2001latent} (see Methods for details), and age groups (see \Cref{fig:appendix_OR_topics,fig:appendix_OR_age_groups} in the Appendix, respectively).
Thus, we confirm that male authors receive negative judgments more frequently compared to female authors, regardless of context and age.
However, it still remains unclear which of the two hypothesized mechanisms leads to this association:
Is this disparity due to a gender bias against male authors?
Or could it be that males are more confident in sharing morally ambiguous stories they are involved in?

\spara{Matching approach for the comparison of similar situations.} 
To test whether judges are biased by the self-disclosed gender of the authors, we use propensity score matching with caliper~\citep{rosenbaum1983central} to compare the moral judgments of pairs of submissions that describe similar situations but have authors of different genders.
The concept of similarity between situations is admittedly blurry.
While the literature has produced several taxonomies to code a situation~\citep{parrigon2017captioning,rauthmann2014situational,tomita2021similarity}, we found them not detailed enough to capture all relevant aspects to assess the similarity between two stories.
Additionally, these taxonomies are difficult to scale by automation.
Therefore, we choose a machine-learning approach to approximate situational similarity using textual embeddings derived from transformer-based Large Language Models (LLMs), which have been shown to outperform other text representations in causal inference~\cite{weld2022adjusting}.

We employ a BERT model~\cite{devlin2018bert} as the propensity scorer, tasked with predicting the self-disclosed gender of protagonists (i.e., the treatment assignment) based on the text of their submissions (see \Cref{app:train_propensity_scorer}). 
To mitigate the model's reliance on gendered words as shortcuts for predicting authors' gender (e.g., the expression ``my wife'' being highly predictive of the protagonist being male), we implement a ``gender neutralization'' strategy~\citep{hall2019name}. 
This strategy consists of swapping all gendered words in a submission with a 50\% probability during training (e.g., ``wife'' changed to ``husband''; see \Cref{app:gender_neutralization} for additional details). 
This approach is crucial because features that predict treatment assignment can be problematic in causal inference~\citep{stuart2010matching}.

Next, we define a distance measure between submissions that considers their semantic similarity and other relevant covariates such as authors' age and the main topic of the submissions (LDA cluster), in addition to the propensity score.
Semantic similarity is obtained via a language model that encodes the content of submissions into semantically meaningful embeddings~\citep{reimers2019sentence}.
Based on this notion of distance, we match submissions by minimizing the pairwise semantic distances between male- and female-authored submissions.
Matches are accepted if (i) the semantic distance is smaller than a threshold, (ii) the propensity score difference is smaller than a caliper value, (iii) the age difference between the authors is at most 5 years, and (iv) the submissions belong to the same topic (see Methods for additional details).
This approach allows us to compare similar stories, thereby mitigating the contribution of 
factors that might correlate the gender of the author with the perceived deviance of the shared situation.
Once the matched pairs have been identified, our estimand of interest is the sample average treatment effect on the treated (SATT)~\citep{keith2020text,imbens2004nonparametric}, defined as:
\[ SATT = \frac{1}{N_T} \sum_{i=1}^{N_T} Y_i\left(M\right) - Y_i\left(F\right) \]
where $N_T$ represents the number of treated submissions (i.e., written by male authors) with a match, $Y_i\left(M\right)$ is the community judgment of the $i$-th treated submission (1 if judged deviant, 0 otherwise), and $Y_i\left(F\right)$ is the outcome of the untreated submission (i.e., written by a female author) that has been matched to the $i$-th treated submission.
In other words, the SATT quantifies how the community judges male authors when they share narratives akin to those shared by female authors.
Unlike the population average treatment effect, SATT estimates the effect on a pruned sample of submissions for which we can find a match satisfying the semantic constraints, and should not be used to infer the effect on the larger population (i.e., the whole subreddit)~\citep{keith2020text,greifer2023choosing,imai2008misunderstandings}.
In other words, SATT estimates the effect in the population of submissions for which male and female protagonists share comparable narratives, in line with the purpose of this study.

\begin{figure}[h]
    \centering
    \includegraphics[width=0.99\textwidth]{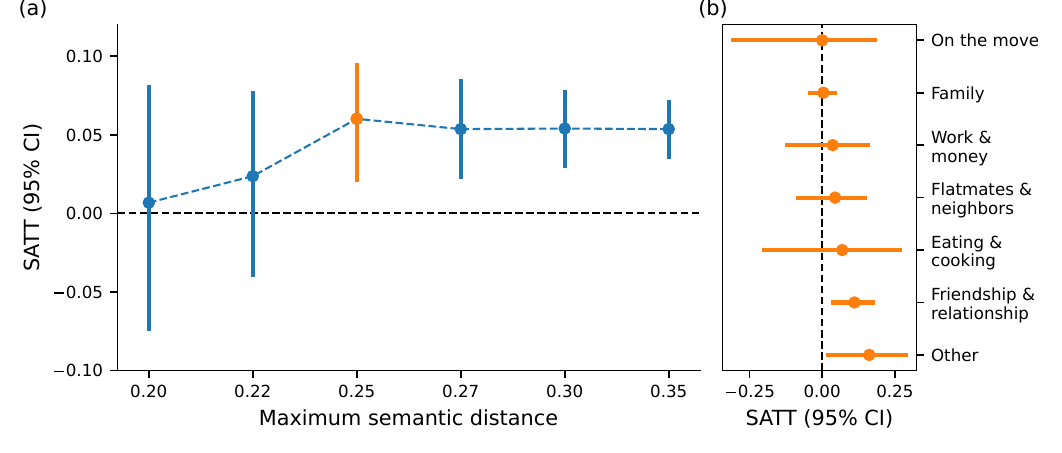}
    \caption{Causal effect of the protagonist's gender on moral judgment. (a) Sample average treatment effect on the treated (SATT) for different values of the maximum semantic distance. 
    We report the SATT from maximum semantic distance equal to 0.20 because of the small amount of matches below this value ($N<120$). 
    (b) SATT for each topic corresponding to the matches obtained with maximum semantic distance equal to 0.25.
    Vertical bars correspond to bootstrap 95\% confidence intervals.
    }
    \label{fig:att_vs_max_distances}
\end{figure}

\spara{Situations explain away the difference.}
Our analysis reveals weak to no evidence supporting a gender bias in community judgments when the narratives are similar.
\Cref{fig:att_vs_max_distances}a shows how the SATT varies for different values of the maximum semantic distance.
The effect of gender is small and not statistically significant for small values of the maximum semantic distance.
However, as the maximum semantic distance increases, a small yet significant effect emerges (SATT $= 0.06$, $95\%$ CI: $0.02$ -- $0.10$) in a matched sample composed of \num{699} pairs of submissions ($6\%$ of submissions authored by male protagonists).
For larger maximum matching distances, the SATT does not increase further.

To contextualize the SATT obtained at a maximum semantic distance of 0.25, we repeat the initial association test on the subset of matched submissions.
Indeed, matching can be used as a preprocessing approach to clean the data by selecting a subset of items such that treatment and control units are similar~\citep{ho2007matching}.
By doing so, we can assess how the association differs from the whole dataset.
In the matched sample, male protagonists are 1.52 times more likely to receive a negative judgment ($95\%$ CI: $1.15$--$2.01$, Fisher's exact test $p\text{-value} = 0.003$), compared to 2.21 in the whole dataset.
This change indicates that our matching approach has identified a subset of submissions where the association between gender and judgment is $1.45$ times smaller than in the whole dataset, thanks to the selection of narratives shared by both male and female protagonists.
The difference between the two odds ratios is deemed significant according to the Breslow-Day test ($\chi ^ 2 = 7.73$, $p$-value $= 0.005$).

Next, we stratify the SATT across the topics identified by LDA to understand if there is a differential contribution of the topic to the observed SATT.
Similarly to \Cref{fig:att_vs_max_distances}a, we depict in \Cref{fig:att_topics} the SATT for each topic as a function of the maximum semantic distance and in \Cref{fig:att_vs_max_distances}b the values obtained for maximum matching distance equal to 0.25.
For most of the topics, the SATT consistently hovers around zero.
The small sample sizes may hinder the ability to measure small causal effects.
For instance, topics such as ``Eating and cooking'' and ``On the move'' comprise $29$ and $16$ matched pairs, respectively, for a maximum semantic distance of 0.25.
In contrast, topics like ``Friendship and relationship'', ``Family'', and ``Other'' (i.e. a miscellaneous category including submissions that do not belong to other topics) have a larger number of matches.
Two of them (``Friendship and relationship'' and ``Other'') align with the overall pattern of the SATT depicted in \Cref{fig:att_vs_max_distances}, where the SATT is not significantly different from zero for small maximum semantic distances but becomes positive for thresholds of 0.25 or larger.
However, the SATT of the ``Family'' topic is not significantly different from zero despite being the largest topic, suggesting that the small observed gender bias may manifest differently depending on the context.
Overall, the SATT measured on the whole matched sample is mainly driven by one single topic, ``Friendship and relationship''.

\spara{The exceeding bias is due to more distant matches.} 
Our findings illustrate a weak or absent gender bias in moral judgment and rely on textual embeddings derived from Large Language Models to identify pairs of similar submissions.
However, the quality of such pairs may confound our findings.
Indeed, pairs of submissions that do not represent similar situations might carry the crude \textit{male-blame} association observed in the whole dataset, thereby escaping our efforts to control for biases.
Matching methods typically check for covariate balance to assess the quality of the matched sample~\citep{stuart2010matching},
but encoding textual data into high-dimensional spaces may invalidate these diagnostics~\citep{mozer2020matching}.
Nonetheless, unlike other types of data, texts can be easily interpreted by humans, thus allowing for the manual assessment of the quality of the matches by conceptualizing a notion of similarity~\citep{keith2020text,mozer2020matching}, albeit introducing some subjectivity in the task.

To evaluate the quality of matched pairs, we conduct a manual evaluation of the matches obtained with a maximum matching distance of $0.25$.
This threshold represents the most conservative value that produces a significant gender bias in judgments.
Five raters familiar with the \texttt{r/AITA} dataset annotated a random sample of $100$ matches (16\% of the matched submissions), with each match annotated by 3 different annotators (see Methods section for additional details).
We employ a five-points Likert scale, ranging from 1 (``\textit{Very dissimilar}'') to 5 (``\textit{Very similar}''), to evaluate the similarity between the situations described in each pair of submissions.
The inter-annotator agreement, measured by Krippendorff's alpha ($\alpha=0.42$), indicates agreement to some extent among annotators despite the subjective nature of the task.
After aggregating the scores, 63\% of matched pairs received a score of 4 or higher (i.e., ``\textit{Somewhat similar}'' or ``\textit{Very similar}''), while 28\% scored 2 or lower (i.e., ``\textit{Somewhat dissimilar}'' or ``\textit{Very dissimilar}''), with the remainder scoring 3 (i.e., ``\textit{Neither dissimilar nor similar}''). 
The result of the manual evaluation indicates that the situations described in the matched pairs of submissions are similar in most of the cases.
We provide examples of matches submissions for different levels of similarity as assessed by the annotators in \Cref{tab:examples_matches}  and additional analyses in \Cref{app:sec_agency_vs_gender_and_judgment}. 

\begin{figure}[h]
    \centering
    \includegraphics[width=0.99\textwidth]{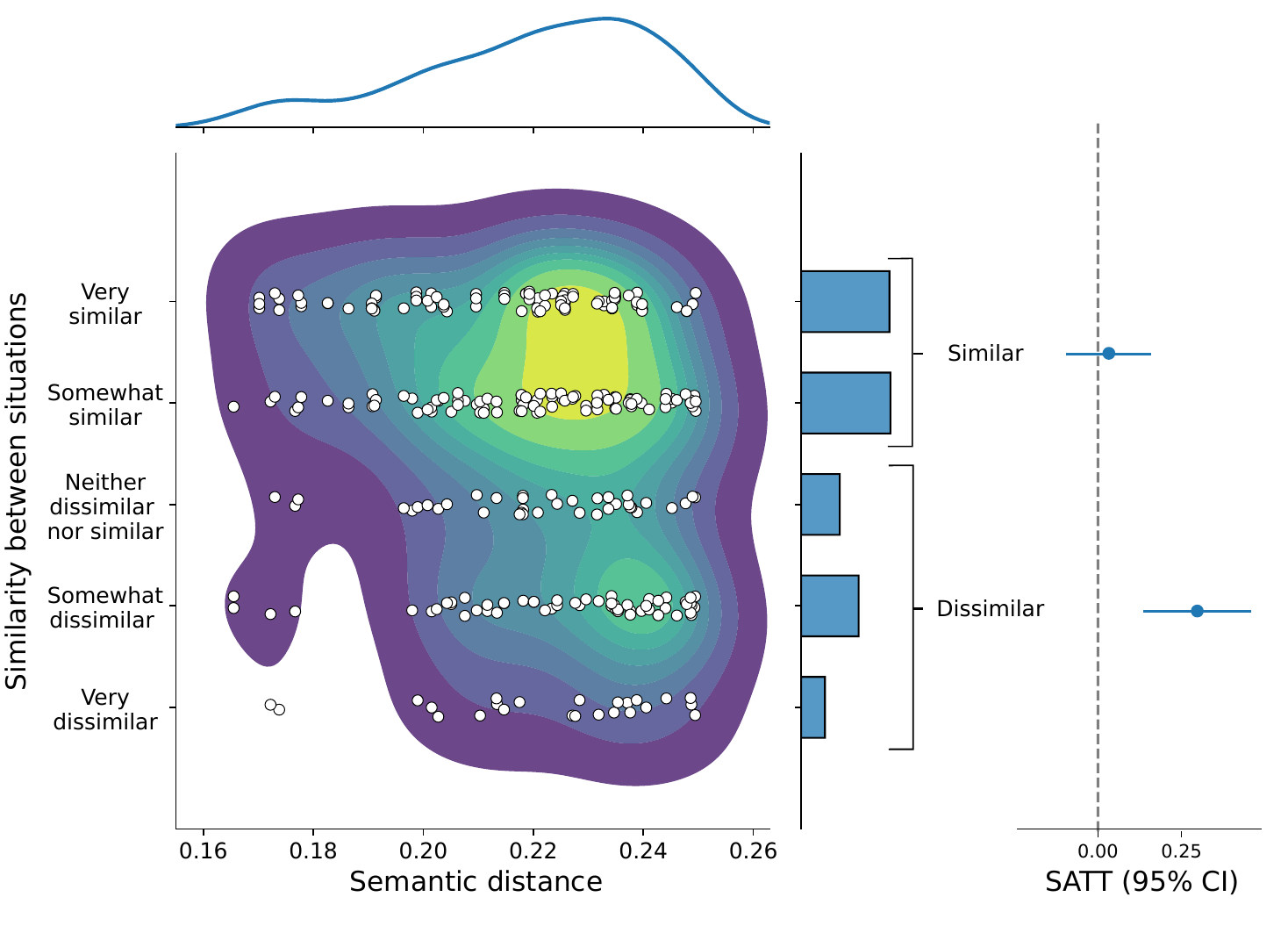}
    \caption{Joint distribution of annotator judgments and distance between matched submissions obtained with maximum matching distance at 0.25. 
    Dots correspond to single annotations of matched submissions (not the aggregated judgment through median aggregation) and the joint plot is obtained through kernel density estimation. 
    Colors range from purple (low density) to yellow (high density). 
    The marginal distributions are shown on top and on the right.
    On the right, SATT and 95\% confidence interval obtained separately for the matched pairs evaluated as similar or dissimilar.
    }
    \label{fig:eval_distance_joint}
\end{figure}

To further understand if the matching procedure effectively reflects the concept of situational similarity, we measure the correlation between the semantic distance and the scores given by annotators.
\Cref{fig:eval_distance_joint} shows the joint distribution of annotators' judgments (disaggregated) and the semantic distance between the matched submissions.
For small distances, the evaluations concentrate around ``\textit{Very similar}'' and ``\textit{Somewhat similar}'' (i.e., 5 and 4 on the Likert scale), indicating that the similarity between situations is effectively captured by a small semantic distance between the submissions.
Instead, dissimilar pairs start to be matched for semantic distance larger than 0.20.
A correlation test between semantic distance and human evaluations finds no evidence for such a relationship (Kendall Tau-b~$= -0.10$, $p\text{-value} = 0.17$), suggesting no significant relationship between the similarity of situations and the semantic distance.
However, the matching procedure is still able to identify effectively similar situations.

Despite the majority of pairs representing similar situations, the presence of a 37\% of dissimilar ones may distort the SATT towards positive values.
To test this intuition, we estimate the SATT for the similar and dissimilar pairs of submissions separately.
As expected, we observe that similar matches produce a SATT that is not statistically different from zero (SATT=0.03, 95\% C.I. -0.10 - 0.16), whereas dissimilar ones result in a positive effect (SATT=0.30, 95\% C.I. 0.14 - 0.46), as shown in \Cref{fig:eval_distance_joint}.
Therefore, the positive and significant gender bias observed for the semantic distance threshold at 0.25 or higher is likely a residual effect due to the presence of dissimilar matched pairs, which contribute positively to the SATT.
Thus, these observations point to a lack of gender bias in moral judgments on \R{AITA}.

\section{Discussion}
In this study, we measure the causal relationship between the self-disclosed gender of users sharing morally ambiguous situations and the moral judgment received by the Reddit community.
Our study design leverages moral judgments from \texttt{/r/AITA}, which has two main advantages.
First, Reddit users are likely to share sensitive stories and honest judgments as their accounts are pseudonymous~\cite{de2014mental,ammari2019self,balsamo2023pursuit}.
Second, the voting mechanism of \texttt{/r/AITA} taps into the wisdom of the crowd and allows a large number of users to express their opinions on the specific situation depicted in the submission.
This, in turn, provides the analysis with a large sample of judgments.
We have found that most of the apparent effect that makes male protagonists more likely to receive a negative judgment~\citep{Candia2022Social,botzer2022analysis,xi2023blame} disappears when controlling for the situation described in the post.
That is, disclosing the gender of the author does not directly affect the judgments received.
This effect survives only in the subset of submissions related to friendship and relationships.

Our results contradict the expected gender bias under the moral typecasting framework~\citep{gray2009moral}.
This framework has important implications in the real world regarding gender biases~\citep{reynolds2020man}, such as the fact that men are more likely seen as perpetrators of moral violations than women because the former are more likely to be seen in roles of power~\citep{arnestad2020hetoo,leiby2021gendered}.
In our case, the situation described instead captures most of the differences in judgments between the genders.

One likely explanation for our result is that males tend to describe `riskier' situations than females.
Indeed, the financial and economic literature has repeatedly observed that ``women are more risk-averse than men''~\citep{croson2009gender}.
Similar results have been found by sociologists and psychologists~\citep{eagly1995science,byrnes1999gender}.
If we assimilate disclosing a personal story that might violate social norms and cause discomfort with risk-taking, then our result is understandable.
Women would self-censor more than men in situations that they consider crossing the line.
Indeed, there is evidence that men are more \emph{overconfident} in their behavior, and this overconfidence translates into a lower return on their investments~\citep{barber2001boys,johnson2006overconfidence}.
A similar phenomenon might be at play in our case: men are overconfident in their understanding of social norms.
This overconfidence causes them firstly to put themselves in  `dubious' situations and secondly to be more likely to share them, thus receiving an overall more negative judgment.
Therefore, the apparent difference in judgments is \emph{not} due to bias from the judges.

A complementary explanation stems from differences in how women and men use online fora.
Broadly speaking, and conscious of the risk of stereotyping, men tend to use online groups primarily for information seeking, while women for encouragement and support while sharing their personal experiences~\citep{klemm1999gender}.
Indeed, women offer more social support on online social networks~\citep{tifferet2020gender}, and engage more often with coping strategies when dealing with distress, e.g.,  through verbal expression of emotions to seek social support~\citep{tamres2002sex}.
Women also more easily provide social support, whereas men (who emphasize achievement, autonomy, and emotional control) have a harder time in seeking and obtaining social support~\citep{barbee1993effects}.
Moreover, men are less likely to provide or seek social support, and when they do engage in social support it is less likely to involve emotion-focused support~\citep{zhou2017ask}.
Males and females may manifest stress differently in their relationships, women tend to perceive/report higher levels of stress, and at the same time, they tend to provide and seek more social support~\citep{kneavel2021relationship}.
Furthermore, social support is more effective for women~\citep{wang2015crosscultural}.
People who express more emotion online receive more support~\citep{yang2018exchanging,li2019communicating}, and women tend to express their emotions more than men in their language~\citep{mulac2006empirical}.
These differences might explain both why males are overconfident and how women use the community as a social support group~\citep{Candia2022Social}.
Indeed, the community is aware of the tendency to use the subreddit for \emph{validation}, and has been actively discussing it.\footnote{\url{https://www.reddit.com/r/AmItheAsshole/wiki/faq/\#wiki_this_subreddit_is_a_validation_circle-jerk._where_are_the_assholes.3F}}

Our findings and the supporting literature should also be considered in light of the emerging trends of AI assistants based on LLMs \cite{muthusamy2023towards,patil2024transformative,baronchelli2024shaping}.
These models are trained on publicly available user-generated data such as Reddit.
As with every algorithmic automation, a chatbot assistant trained on biased data would amplify and reinforce existing behaviors unfair towards specific subpopulations.
Clearly, this can be harmful \emph{per se}.
Still, the scenario might further spiral down as more automatically generated text (trained on biased data) will be used to train novel LLMs with even stronger biases.
Thus, our study highlights additional potential risks in deploying and using LLMs to augment, or even automate mental health support services \cite{stade2024large}.
At the same time, using a causal approach in the machine-learned model might be a viable strategy to remove the apparent bias.

We have found that disclosing the gender of the author does not affect the moral judgment received, so where does the observed \textit{male-blame} association come from?
Based on our causal assumptions depicted in \Cref{fig:causal-dag}, the other causal path from the author's gender to judgment passes through the likelihood of sharing a specific situation based on the authors' gender.
A hint for this mechanism being in place comes from secondary findings derived from the manual annotation, where we observe that male authors are more agentic than female ones (see \Cref{app:sec_agency_vs_gender_and_judgment}).
In other words, this observation points to male protagonists being more responsible for the event than female ones, and possibly for a higher likelihood of males sharing morally ambiguous situations, if we assume that deviant behavior correlates with higher agency.
Testing this mechanism is a promising direction for future work, which requires quantifying the level of agency of protagonists at scale.

While many gender biases remain~\citep{eisenchlas2013gender,elinder2012gender,kreager2016double}, our result contributes to a body of hopeful results that show their absence~\citep{greve-poulsen2023no}.
We may speculate that a generational change affects these biases, as ``Gen Z'', which composes the primary user base of Reddit, is known to be more sensitive to gender issues~\citep{low2023riskinformed}.
Overall, our results have important implications for gender studies, the study of social norms, and the understanding of socio-technical systems.

\spara{Limitations.}
As with any empirical study, our work comes with its limitations.
We wish to emphasize that our result pertains specifically to the community under study.
This fact has several implications.

First, it only applies to online interactions on a forum mediated via text.
In face-to-face interactions, the situation might be different due to a plethora of other factors such as verbal and body language, visual and social cues, and interactive dynamics.

Second, among online communities, Reddit provides specific affordances that might influence the social processes on it (e.g., echo chambers are less likely to appear on a forum than a social network~\citep{cinelli2021echo,mok2023echo}).
The specific demographic composition of the \texttt{/r/AITA} community should be also taken into account, as its user base is particularly young, female, and US-dominated.\footnote{\url{https://imgur.com/a/POhgZsh} (accessed on 20/03/2024).}

Third, there might be some unobserved confounders that hinder our causal analysis.
For instance, gender homophily might play a role in the determination of judgments (e.g., male judges being more lenient towards male authors).
The demographic characteristics of the judges are not readily available, but it is possible to make an educated guess from their posting habits~\citep{waller2021quantifying}, and this is a promising line of inquiry for future work.
Nevertheless, the final judgment is determined by upvotes, which are completely anonymous.
This limitation can only be overcome via an experimental study with volunteers.

\section{Methods}

\spara{The AITA dataset.} 
\R{AITA} is a subreddit where users submit real-life experiences to seek feedback on their actions from the community.
Community members reply to submissions and eventually vote on the protagonist's behavior by using judgment tags.
These tags---\textit{Not The Asshole} (NTA), \textit{No Assholes Here} (NAH), \textit{You're The Asshole} (YTA), and \textit{Everyone Sucks Here} (ESH)---are commonly used in the community and listed in the voting guidelines of the subreddit.
In addition, community members can upvote comments they agree with (while downvotes are reserved for off-topic discussions or spam).
The difference between upvotes and downvotes is shown in Reddit comments and is known as the ``score'' of the comment.
\R{AITA} uses a bot to automatically assign a final judgment based on the judgment tag contained in the highest-scoring comment 18 hours after the submission is posted.

We collected all submissions posted on \R{AITA} between 2014 and 2020, retrieving all the comments containing one of the judgment tags from the Pushshift Reddit data collection~\citep{Baumgartner_Zannettou_Keegan_Squire_Blackburn_2020}.
Since we are interested in the community judgment towards the protagonist of the submission, we merge \texttt{YTA} with the \texttt{ESH} tag, and the \texttt{NTA} with the \texttt{NAH} tag, as they convey the same judgment towards the protagonist.
These new tags are named \texttt{AH} and \texttt{N\_AH} respectively.
After discarding content authored by bot accounts, the AITA dataset is composed of \num{250770} submissions and \num{6891476} comments (see \Cref{appendix:bot,appendix:data_collect} for additional details about the data collection).

\spara{Identifying the main topics.}
The submissions in the AITA dataset cover a wide range of subjects such as stories about parenting, work, or issues related to living with other people, just to mention a few.
We identify the main topics to achieve two goals: (i) to obtain a coarse-grained clustering of the submissions for exploring the dataset, and (ii) to use these topics in the matching algorithm.
In addition, the division of submissions into topics allows us to assess the consistency of the observed bias across different subjects.
We train a Latent Dirichlet Allocation (LDA) topic model~\citep{blei2001latent} on all the submissions to estimate the probability that a submission belongs to each of $N_T=6$ topics.
The number of topics is chosen by minimizing the perplexity on a held-out set of submissions in a 5-fold cross-validation setting (see \Cref{fig:app_lda_perplexity_training} and \Cref{appendix:topic_detection} for additional details).
Then, we assign each submission to the topic with the highest probability, provided that this probability is higher than the threshold $0.4$.
Submissions not meeting this threshold are assigned to a topic called ``Other'', which includes submissions whose topic is either not well identified or a combination of the other topics.

\spara{Extraction of judgments and demographics.} 
To determine the community judgment for each submission we extract judgment tags from comments (see \Cref{appendix:data_collect} for additional details).
We then weight the judgment tags under each submission by the comment's score (i.e., the difference between upvotes and downvotes) and assign the judgment corresponding to the tag with the largest weighted score to the submission.
This approach considers all the comments that express a judgment, including all users who upvoted such comments.
We discard submissions whose weight is lower than 10 to exclude submissions with few judgments.

To infer the author's gender and age, we extract specific demographic tags commonly employed by users to provide more context to the community.
For instance, ``F26'' denotes that the author identifies as a female aged 26, whereas ``22 M'' refers to a male aged 22.
We use a regular expression to extract this information from the AITA dataset if such tags occur in the proximity of first-person singular pronouns, which indicates that the tags refer to the author of the submission (see \Cref{appendix:OP_demog} for additional details).
We identify the age and gender of the author in $15.6\%$ of the submissions.
We also verify that topics have a uniform distribution of submissions with available demographics to confirm that the self-disclosure of gender does not depend on the topic (see \Cref{fig:disclosure_representativity_across_topic}).

\spara{Propensity score matching.} 
The observed correlation between the received moral judgment and the self-disclosed gender may be biased by other factors.
To measure the direct effect of gender on moral judgments, we make assumptions about the possible causes that lead the community to judge a submission, summarized in the causal graph shown in \Cref{fig:causal-dag}.
The gender and age of the protagonist can affect both the type of situations they can experience (i.e., the arrow from ``Gender'' and ``Age'' to ``Experiencing a situation'') and the likelihood of sharing it on \texttt{r/AITA} (i.e., the arrow from ``Gender'' and ``Age'' to ``Posting on \texttt{r/AITA}'').
For instance, users might not be equally likely to tell morally ambiguous stories.
As suggested by~\citet{Candia2022Social}, male authors may feel more comfortable in sharing more controversial situations, whereas female authors may be more likely to seek for validation by sharing stories where they clearly do not deserve blame.
This effect 
could introduce a reporting bias affecting the association between gender and judgment.
In addition, the age and gender of the protagonist can influence the spectrum of situations they experience
(i.e., the demographic of a person affects the likelihood of experiencing specific situations).
For instance, pregnancy-related situations experienced by women, or work-related situations less likely to be experienced by teenagers.
The judges deliberate based on the submission they read (i.e., the arrow from ``Posting on \texttt{r/AITA}'' to ``Judgment'') and the demographic information of the author (i.e., the arrow from ``Gender'' and ``Age'' to ``Judgment'').
The latter causal link is the effect we aim to estimate by controlling for the context of the stories told by their protagonists.

This conceptualization reflects two assumptions needed to clarify which are the observable and unobservable confounders.
First, we consider the judgment of the community to be derived solely from the information accessible in the text of the submission.
We believe this to be a reasonable assumption since judges have no additional information to consider.
In other words, only the content of the submission affects the moral judgment of the community, which includes the demographic attributes disclosed by the protagonist.
Second, we assume all submissions are judged by a homogeneous jury.
That is, the set of judges on any given story is a random sample from the population of judges.
It follows that the moral judgment does not depend on the specific set of users who contribute to the judgment of the submission.
An intrinsic limitation of Reddit requires this assumption, as the platform does not provide access to the identities of the users who upvote comments, thus making it impossible to identify the judges.

We employ propensity score matching to reduce biases in the estimation of the causal effect~\citep{stuart2010matching}.
To define the distance between two submissions, we use a combination of coarse matching and semantic distance matching within propensity score calipers.
Specifically, we use a document embedding model, SBERT~\citep{reimers2019sentence}, to encode submissions into a semantically meaningful vector and compute document pairwise similarity through cosine distance.\footnote{Model card of the pre-trained model used: \url{https://huggingface.co/sentence-transformers/all-mpnet-base-v2} Even though SBERT is designed for sentence embeddings, this particular model was trained on a dataset that includes a relevant fraction of data from Reddit.}
Then, after having trained the propensity scorer (see \Cref{app:train_propensity_scorer}), we define the distance between two submissions $i$ and $j$ as:
\[ D_{i, j} = 
    \begin{cases}
      1 - \cos\left(\vec{i}, \vec{j} \right), & \text{if} \; \left\vert e\left( i \right) - e\left( j \right) \right\vert < c \\
      & \wedge \; 1 - \cos\left(\vec{i}, \vec{j} \right) \leq D_{\max} \\
      & \wedge \; \tau_i = \tau_j \\
      & \wedge \; \left\vert age\left(i\right) - age\left(j\right) \right\vert \leq \delta \\
      & \\
      + \infty, & \text{otherwise}
    \end{cases}
\]
where $\cos\left(\vec{i}, \vec{j}\right)$ is the cosine similarity between the submission embeddings of $i$ and $j$ obtained via SBERT, $e\left(i\right)$ is the logit of the propensity score of submission $i$, $\tau_i$ is the topic of submission $i$, $c$ is the caliper, $D_{\max}$ is the maximum matching distance, $age\left(i\right)$ is the age of the author of submission $i$, and $\delta=5$ is the maximum difference in age between authors of submissions. 
The caliper is determined as $c = 0.2 \times \sqrt{\frac{\sigma^2_T + \sigma^2_U}{N_T + N_U - 2}}$, where $\sigma^2_T$ ($\sigma^2_U$) is the variance of the distribution of the logit of the propensity score of treated (control) submissions~\citep{wang2013optimal}.
Once all pairwise distances between treated and untreated submissions are computed, we obtain 1:1 matches without replacement via the minimum weight matching algorithm, which minimizes the sum of the distances between the matched submissions.
We opt for 1:1 instead of 1:many matching because it simplifies the manual evaluation.
We consider different values of $D_{\max}$ from 0.15 to 0.35.

We evaluate the matches by assessing the covariate balance through two measures~\citep{stuart2010matching}: standardized difference of means of the propensity score and the ratio of the variances of the propensity score in the treated and control groups.
All the configurations displayed in \Cref{fig:att_vs_max_distances} results in a satisfactory covariate balance according to the two measures.
Specifically, the standardized difference of means results smaller than 0.25 and the ratio of the variances between 0.5 and 2.0.
However, despite the commonplace use of these measures in the literature, texts are embedded in high-dimensional vector spaces, which might hinder finding similar documents across all the dimensions.
Another valid balance check consists of a manual evaluation of the similarity of the matched pairs of submissions, given the ease of interpretation of textual data~\citep{mozer2020matching}.

\spara{Evaluating the quality of matched submissions.}
We employ a manual annotation process to evaluate the quality of the matched submissions obtained with maximum semantic distance of $0.25$.
Five raters, familiar with the data, annotated a random sample of 100 matched pairs of submissions for the evaluation (14\% of the total matched pairs), with each pair evaluated by three different annotators.

The annotation consists of three steps for each pair of submissions. 
In the first and third steps, the annotator is asked to assess the level of agency of the two protagonists.
After reading the title and body of each submission, the annotator is asked to reply to the question ``\textit{How much of the whole event in the text is caused/initiated by the author?}''~\cite{tomita2021similarity} on a 5-point Likert scale ranging from \textit{Not caused by the author} to \textit{All caused by the author}.
The interest in this information is related to the second hypothesis about the observed association between gender and judgment.
That is, if male protagonists share more morally ambiguous situations than female ones, we should expect male protagonists to express a higher level of agency.

In the second step, the annotator is asked to evaluate the similarity between the two stories on a 5-point Likert scale ranging from \textit{Very dissimilar} to \textit{Very similar}.
Since the concept of similarity is not well defined, the annotators refined their common understanding of similarity by discussing it in a preliminary stage in which they annotated a small set of pairs.
Our operationalization of similarity accounts for different narrative elements such as individuals involved in the story (e.g., family members, friends, colleagues), the nature of the actions, the roles of each participant, and the setting when relevant.
For example, a story in which the protagonist yells at a family member is deemed dissimilar from one where a family member yells at the protagonist: despite the individuals involved and the action being the same, the roles are reverted.
As another example, consider a pair of stories consisting of one where the protagonist yells at a friend at home while the other where the protagonist yells at a friend in a public place like a restaurant.
Here, the similarity of characters and actions is outweighed by the difference in places where the action unfolds, so the stories are considered dissimilar.

We ask annotators to evaluate the similarity of the two stories between the assessment of each submission for two main reasons. 
First, since the submissions can be quite long and convoluted, we prefer to minimize the interval between reading the two stories to increase the recall of details.
Second, we wish to avoid the judgment of the similarity between the two stories to be influenced by the annotations applied to each story separately.

To make a final assessment of the quality of the matched submissions, we employ median aggregation to the ratings on the similarity between the pairs.
That is, given the annotations from the three raters to each pair of submissions, we discard the two most extreme judgments.
In cases where the aggregation resulted in ``Neither dissimilar nor similar'', the authors of this study engaged in conflict resolution (12 out of the 100 evaluated matches).
This process led to 2 pairs becoming ``Somewhat similar", 3 pairs becoming ``Somewhat dissimilar", and 7 pairs remaining ``Neither dissimilar nor similar" (i.e., there was no agreement after the conflict resolution). 
This evaluation resulted in $29\%$ ``Very similar", $32\%$ ``Somewhat similar", $14\% $``Neither dissimilar nor similar", and the remaining $25\%$ ``Somewhat dissimilar". 
Overall, $61\%$ of the submissions were rated as ``Somewhat similar" or higher.

\section*{Acknowledgements}
We wish to thank Corrado Monti and Francesco Bonchi for their help on an early version of this work.
LB acknowledges support from Intesa Sanpaolo Innovation Center. The funder had no role in study design, data collection and analysis, decision to publish, or preparation of the manuscript.

\clearpage
\appendix
\renewcommand\thefigure{\thesection.\arabic{figure}} 
\setcounter{figure}{0}  

\renewcommand\thetable{\thesection.\arabic{table}} 
\setcounter{table}{0}

\section{Dataset creation}

\subsection{Bot detection}\label{appendix:bot}

We discard content published by bot accounts.\footnote{Two bot users are used by \texttt{/r/AITA} moderators to perform automatic tasks: \texttt{AutoModerator} and \texttt{Judgement\_Bot\_AITA}. We consider them as bots as well.}
We refer to the subreddit \texttt{r/BotDefense}, where users report other users exhibiting suspicious behavior and the subreddit assigns a flag to the reported user if it meets certain criteria. 
We collect all the submissions in this subreddit and classify as bots all the users for which the service assigned the flair ``banned'', ``declined'', or ``service''.
This amounts to \num{2604}, \num{429}, and \num{98} users respectively.
We discard all the contents produced by this set of users (i.e., both submissions and comments).

\subsection{Data collection}\label{appendix:data_collect}

We collect all the submissions and comments on \texttt{/r/AITA} from the beginning of 2014 to the end of 2020 from the Pushshift Reddit data collection~\citep{Baumgartner_Zannettou_Keegan_Squire_Blackburn_2020}. 
We filter out all submissions whose title does not start with ``AITA'' (i.e., ``Am I the asshole'') or ``WIBTA'' (i.e., ``Would I be the asshole''), since authors who want to receive community judgment must start the tile with such tags according to community guidelines.
In addition, we discard submissions whose final judgment is ``NFO'', indicating that the author did not provide enough information to let the community deliberate.
Submissions whose body was deleted or removed are discarded as well.
Then, we collect the comments under the filtered submissions containing at least one judgment tag.
After the removal of contents published by bot accounts, the dataset contains \num{252269} submissions and \num{8191812} comments.

In the next step, we extract the judgment tags from the comments.
Users do not always use judgment tags only to express their judgment.
For example, they can discuss the behavior of the author and use judgment tags as abbreviations, instead of using them with intent of expressing their judgment.
In addition, the ``NAH'' tag may be confused with the informal spelling of ``no'', since users use tags with various case variants (e.g., ``NAH'' as well as ``Nah'' and ``nah'').
To extract judgment tags from comments whose intent is to cast a judgment on the behavior of the author, we develop a set of rules by considering where the tag occurs within the comment:
\begin{enumerate}
    \item The tag is the only word in a line, irrespective of the case. 
    In this case, the tag is separated from the comment body and we can safely assume that the judge intends to use it to express their vote. 
    This is the only case in which we considered the tag even in the presence of other tags in the comment.
    \item The tag is the only word of a sentence, irrespective of the case.
    Similarly to the previous rule, the author chooses to isolate the tag from the body of the comment. 
    In this case, we do not consider the tag ``Nah'' and ``nah'' because of the ambiguity with the informal spelling of the word ``no''. 
    However, this happens only in approximately 8k comments.
    \item The tag is the first word of a line and written in upper case, or it is followed by some special characters irrespective of the case.
    In this scenario, the upper case or the presence of a special character isolates the tag from the rest of the comment, similar to the previous cases.    
    The special characters we consider are: ``.'', ``-'', ``(space)-'', ``;'', ``:'', ``(space):'', ``(double space)''.
    \item The tag is written in upper case in a sentence composed of at most 6 words. 
    This rule covers a non-negligible fraction of cases in which the judge wants to emphasize their judgment. 
    For example, ``OP, you are clearly NTA!'', ``Uh yeah, YTA'', and ``I think ESH''. 
    As an exception to this rule, we discard all cases where the short sentence contains the word ``if'' or ends with a question mark, as in these two cases the sentence may be hypothetical or a question.
\end{enumerate}

Each rule is responsible for selecting $17\%$, $31\%$, $32\%$, and $5\%$ of comments, respectively, while the remaining $15\%$ of comments are discarded.
Approximately half of the discarded comments are removed because they have a negative score (difference between upvotes and downvotes).
We decided to remove them as they are likely to violate the rules of the community.
Finally, after having removed submissions with no evaluations, our dataset comprises \num{250770} submissions and \num{6891476} judgments.

\subsection{Topic detection}\label{appendix:topic_detection}

To get a first coarse division of submissions into topics, we employ a topic modeling technique, LDA~\cite{blei2001latent}, to assign a distribution of topics to each submission.
We use the text of the submissions as input for LDA after a preprocessing step, described next.
First, we discard submissions that are either too long (more than \num{3000} words, according to subreddit length rules) or too short (less than 100 words, corresponding to the $5$-th percentile of the distribution of submissions' word length), which encompasses $9.8\%$ of submissions overall.
Then, we lemmatize and remove stopwords from the submissions.
We discard lemmas occurring in more than $50\%$ and in fewer than $10$ submissions since those are not informative for topic attribution as either too frequent or too rare.
These preprocessing steps result in \num{226229} submissions used to train the LDA topic modeling.

\begin{figure}
    \centering
    \includegraphics[width=0.99\textwidth]{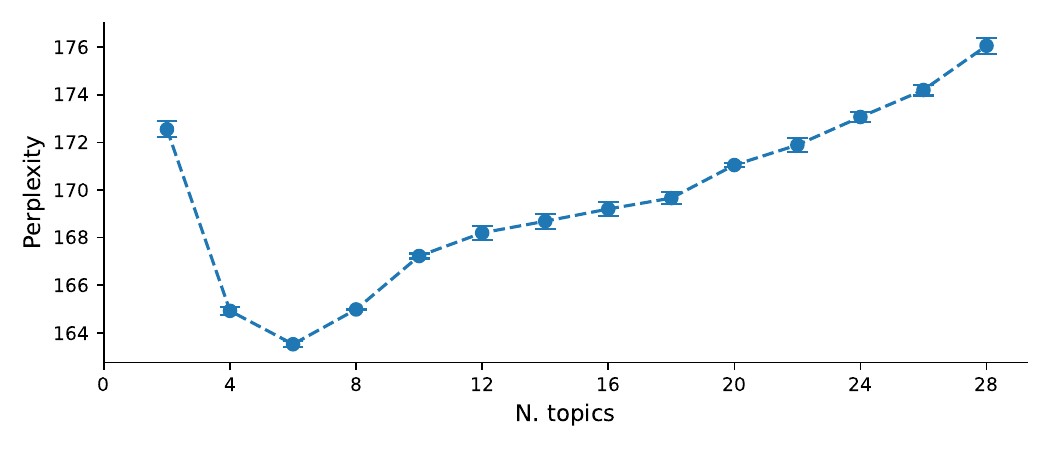}
    \caption{Perplexity on held-out set. Circles indicate the mean and vertical bars indicate the standard error of the mean of the perplexity obtained in a 5-fold cross-validation.}
    \label{fig:app_lda_perplexity_training}
\end{figure}

We use the LDA implementation by the Gensim library~\citep{rehurek_lrec} with a number of topics ranging from 2 to 30.
We select the optimal number of topics by minimizing the perplexity computed in a 5-fold cross-validation setting.
As shown in \Cref{fig:app_lda_perplexity_training}, the perplexity is minimized with six of topics.
We then train LDA on the whole set of documents with the optimal number of topics to obtain a topic distribution for each submission.
\Cref{tab:lda_topics} shows the 15 most important words for each topic.
We interpret and give a title to each topic by looking at this table and reading the 5 submissions with the highest probability for each topic.

\begin{table}
	\centering
	\caption{Top 10 words for the identified topics.}
	\begin{tabular}{p{3cm} p{9cm}}
	\toprule
		Topic & Top 10 words \\ 
		\midrule
		Eating and cooking  & eat, food, wear, buy, thing, dinner, hair, drink, cook, feel \\ \cmidrule(lr){1-2}
		On the move & play, car, game, drive, walk, start, minute, leave, phone, watch       \\ \cmidrule(lr){1-2}
		\makecell[l]{Flatmates and\\neighbors} & dog, room, house, live, home, leave, sleep, day, clean, roommate \\ \cmidrule(lr){1-2}
		\makecell[l]{Friendship and\\relationship} & friend, feel, talk, thing, good, people, start, year, day, girl \\ \cmidrule(lr){1-2}
		Family & mom, family, dad, sister, year, parent, brother, kid, mother, husband \\ \cmidrule(lr){1-2}
		Work and money & work, pay, money, job, day, week, year, month, buy, school \\
	\bottomrule
	\end{tabular}
	\label{tab:lda_topics}
\end{table}

\subsection{Extraction of author demographics}\label{appendix:OP_demog}

We extract the demographic information of the authors from their submissions.
Authors often use demographic tags to provide more context to their stories.
We develop a regular expression to extract tags that contain information about the author's age and gender.
The regular expression selects demographic tags in the proximity of first-person singular pronouns (e.g., ``I am a F26 [...]'', ``My (F26) [...]'') and handles variations of the patterns such that the gender tag can be before or after the age, and possibly separated by a whitespace (e.g., ``F 26'', ``26 F'', or ``26F''). 
In total, \num{39005} submissions contain this demographic tag, accounting for $15.6\%$ of the submissions. 

To account for other gender identities, we modify the regular expression to find strings containing a two-digit number (representing the age) followed or preceded by up to four capital letters, which might potentially refer to other gender tags.
We manually inspect tags occurring more than 2 times, and find acronyms related to non-binary genders (NB) and transgender (MTF or FTM, meaning users transitioning from male to female or female to male respectively), just to name a few.
After including also case-insensitive versions (e.g., MtF) and variations (e.g., M2F), we identified 309 additional submissions with other gender tags. 
Since these constitute less than 1\% of the total users with available demographic information, we consider only the binary categorization of gender (M and F) due to the limited data on other gender identities.

\subsection{Basic statistics of the dataset}

\begin{figure}
    \centering
    \includegraphics[width=1.\textwidth]{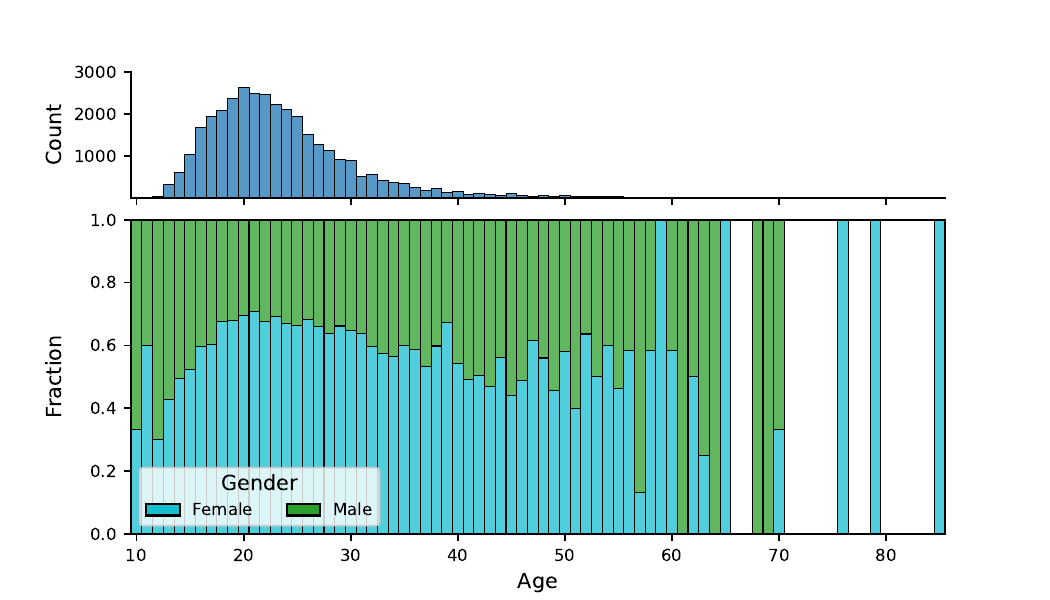}
    \caption{Fraction of male and female authors for different ages. 
    The top plot shows the aggregate distribution of authors' age.
    }
    \label{fig:app_distrib_age_gender}
\end{figure}

The final dataset consists of the submissions for which we are able to extract the judgment tag and the demographic information of the protagonist, which comprises \num{35375} submissions. 
After discarding submissions with fewer than 10 judgments, the final dataset has \num{33421} submissions.
Protagonists have a median age of 22 years (IQR: 19-26), which is homogeneous across male and female protagonists.
\Cref{fig:app_distrib_age_gender} shows the distribution of protagonists' age, stratified by gender.

\begin{figure}
    \centering
    \includegraphics[width=0.99\textwidth]{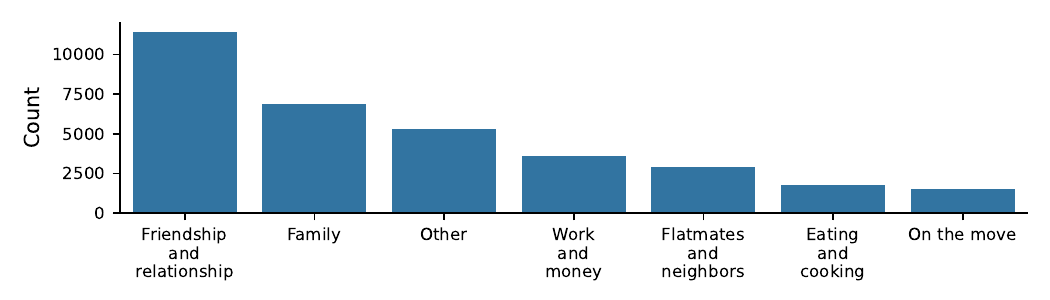}
    \caption{Distribution of topics in the AITA dataset as returned by the LDA topic model. Submissions are assigned to the topic with highest probability if higher than $t_{topic}=0.40$. Otherwise, they are assigned to the ``Other'' topic.}
    \label{fig:app_distribution_of_topics}
\end{figure}

The most frequent topic is about friendship and relationships, followed by family, as shown in \Cref{fig:app_distribution_of_topics}.
These two topics account for $55\%$ of submissions.

\begin{figure}
    \centering
    \includegraphics[width=0.99\textwidth]{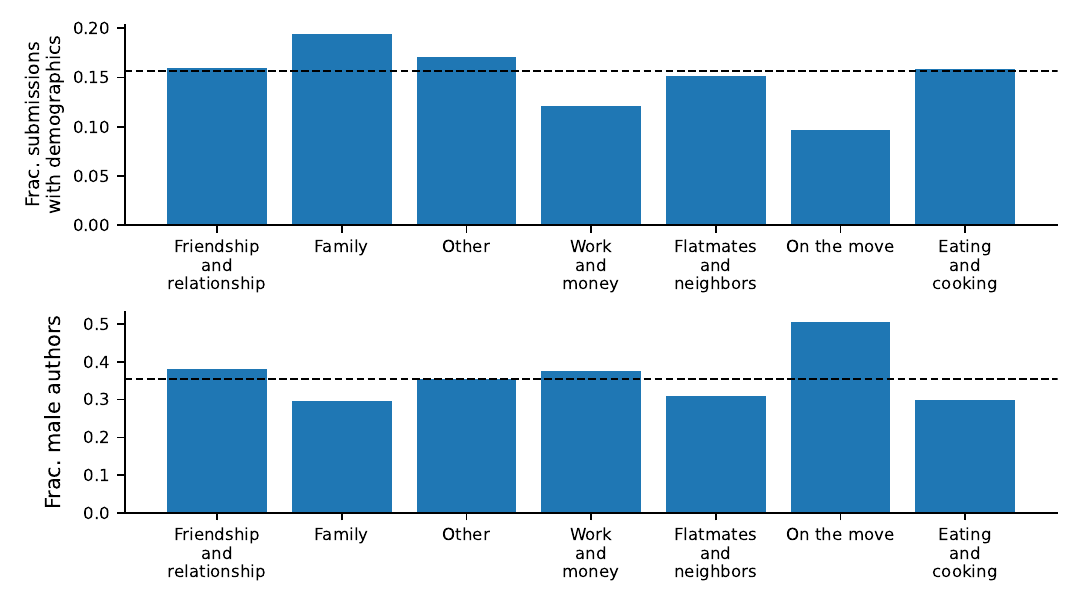}
    \caption{Distribution of submissions with available demographics (top) and male authors (bottom) for each topic identified by LDA.
    The black dashed line refers to the fraction of submission from which it has been possible to extract the demographic tag (top) and the fraction of male authors (bottom).
    }
    \label{fig:disclosure_representativity_across_topic}
\end{figure}

Since the dataset results from the availability of different information, we verify to what extent the fraction of submissions with available demographic information is uniform across topics.
The top panel of \Cref{fig:disclosure_representativity_across_topic} shows that for most of the topics, the fraction of submissions with available demographics ranges from $15\%$ to $17\%$. 
The topics whose fractions differ the most are ``On the move'' ($10\%$), ``Work and money'' ($12\%$), and ``Family'' ($19\%$).
This result indicates that authors self-disclose their demographic information evenly across topics.
The fraction of male authors is also homogeneous across topics, as shown in the bottom panel of \Cref{fig:disclosure_representativity_across_topic}, except for the topic ``On the move''.

\begin{figure}
    \centering
    \includegraphics[width=0.99\textwidth]{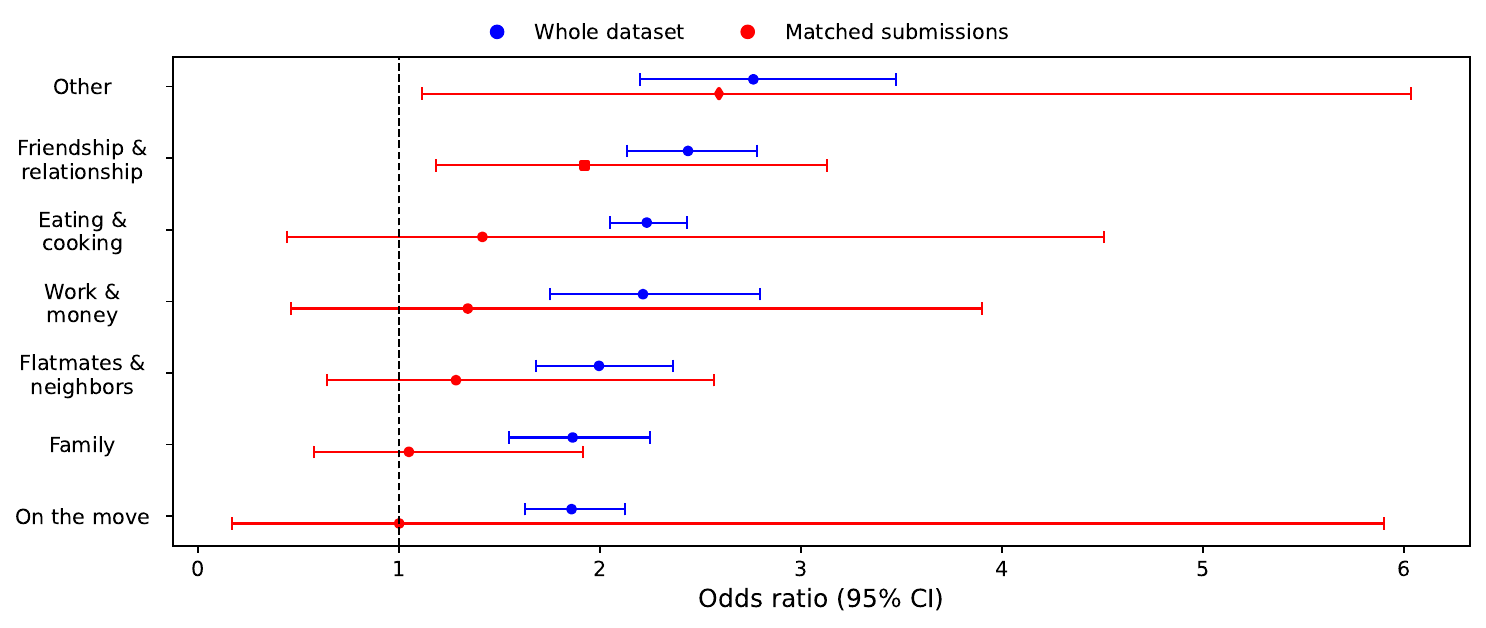}
    \caption{Odds ratio of receiving a negative judgment given the gender of the author for each topic separately. Results show separately the associations for the whole dataset (blue lines) and the matched sample (red lines). Horizontal bars refer to 95\% confidence intervals. Empty markers indicate lack of significance at 0.05 level. \\ 
    $\diamondsuit$ : $p < 0.05$,
    $\Box$ : $p < 0.01$, 
    $\bigcirc$ : $p < 0.001$ }
    \label{fig:appendix_OR_topics}
\end{figure}

\begin{figure}
    \centering
    \includegraphics[width=0.99\textwidth]{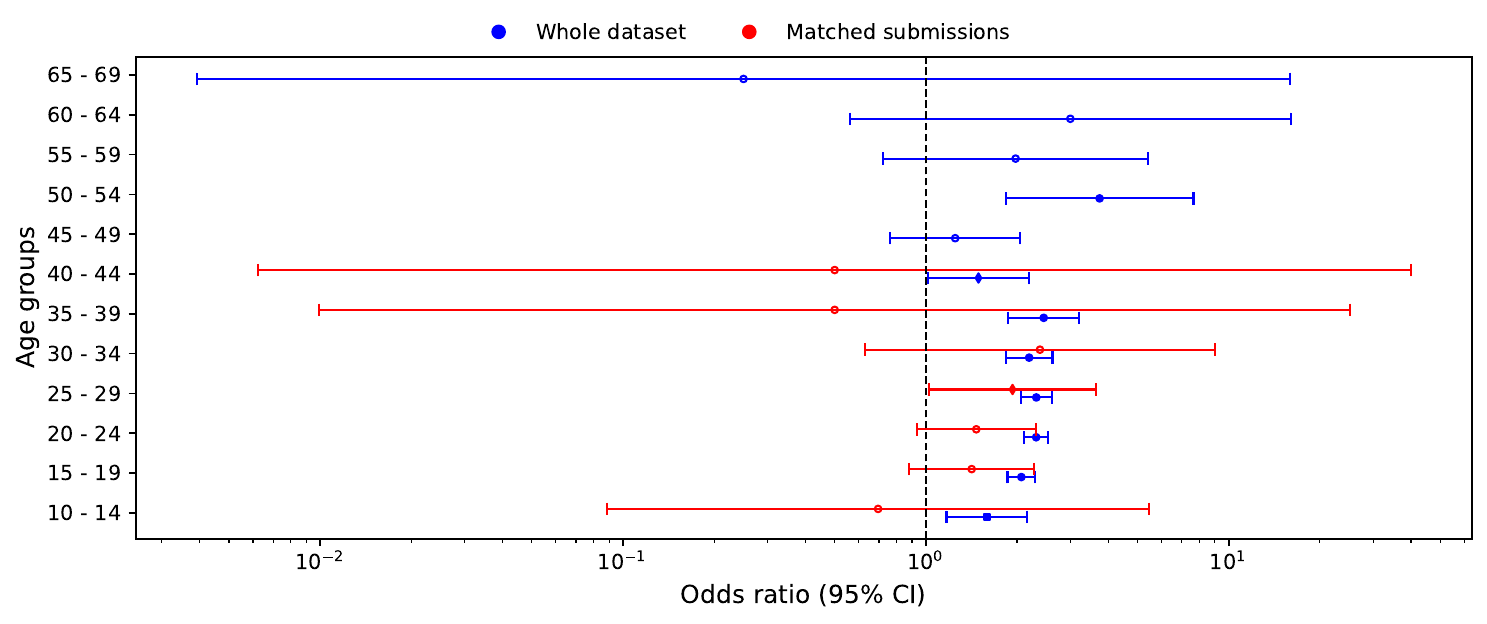}
    \caption{Odds ratio of receiving a negative judgment given the gender of the author for different age groups. Horizontal bars refer to 95\% confidence intervals. Empty markers indicate lack of significance at 0.05 level. \\ 
    $\diamondsuit$ : $p < 0.05$,
    $\Box$ : $p < 0.01$, 
    $\bigcirc$ : $p < 0.001$ }
    \label{fig:appendix_OR_age_groups}
\end{figure}

\section{Estimation of gender bias with propensity score matching}

\subsection{Training the propensity scorer}
\label{app:train_propensity_scorer}

To train the propensity score model, which should predict the ``treatment assignment of a submission'' (i.e., the self-disclosed gender of its author), we train a BERT model (\url{https://huggingface.co/google-bert/bert-base-uncased}) to predict the self-disclosed gender of the protagonist.
The input to the model consists of the concatenation of the title and text of the submission.
Since the gender of the protagonist is extracted from the demographic tags present in the text, we remove all the demographic tags from the text to prevent information leakage.

We train the propensity scorer for three epochs using a learning rate of $2 \times 10^{-5}$, batch size $64$, weight decay  $0.01$, linear warm-up for $10\%$ of training steps, and class weighting reflecting the relative fraction of male- and female-authored submissions in the training set.
We employ an early stopping strategy on $10\%$ of the training set.
Such parameters were chosen after a 5-fold cross-validation on the learning rate ($2 \times 10^{-5}$, $5 \times 10^{-5}$) and the number of epochs ($3$, $5$), which resulted in negligible differences.

\subsection{Gender neutralization strategy}
\label{app:gender_neutralization}

The propensity score estimates the probability of a submission to be treated (i.e., written by a male protagonist) from its textual content.
In other words, it estimates the ``gender typicality'' of a specific situation from the dataset.
This allows the matching procedure to match pairs of situations having similar gender typicality, although the gender of their authors is different.
At the same time, it prevents matching pairs of submissions whose gender typicality is extreme, as a situation that is extremely typical for only one gender is unlikely to be experienced by someone of the other gender.

Analyzing the results, we noticed that our first propensity scorer (``base'' model) relies on the presence of gendered words in the text to predict the gender of the author instead of focusing on the gender typicality of the situation.
Let us consider the two examples reported in \Cref{tab:examples_gender_neutralization}.
The first example is about pregnancy and the propensity score returned by the base model is 3\%, meaning that the author of the submission is very likely a woman.
If we swap all gendered words in the text, the predicted propensity score remains equal to 3\% (see the ``Gender swap'' column under the base model in \Cref{tab:examples_gender_neutralization}).
Thus, the base propensity scorer is not sensitive to the changed words.
The second example has a propensity score of 96\%, indicating that the author is very likely a male according to the model.
Differently from the previous example, by replacing the only word ``wife'' with ``husband'', the propensity score changes dramatically to 5\%.
This large change in the propensity score is likely the result of spurious correlations that relate certain gendered words to the gender of the author of the text.
In this specific case, ``my wife'' indicates that according to the model the author has to be a male likely because in most of the submissions of the AITA dataset, the marriage relationship holds between a man and a woman.
Given our goal of pairing similar situations that differ in the author's gender, the occurrence of specific words correlating strongly with the authors' gender poses a problem for our methodology.
Indeed, the strong difference in the propensity score of such a pair of submissions, one with the word ``wife'' and the other one with the word ``husband'', would prevent them from being matched, even though they are a hypothetically perfect match.

\begin{table}[h!]
	\centering
	\caption{Effect of gendered words on the propensity score. 
    The table shows the propensity score of submissions obtained from the base and the gender-neutralized models.
    The ``Original text'' column refers to the propensity score of the original text, while the ``Gender swap'' column refers to the text where gendered words were swapped.
    The swapped words are underlined and in bold.
    We rephrase and summarize the content of the submissions to ensure anonymity.
    }
	\begin{tabular}{p{5.5cm} p{1.5cm} p{1.5cm} p{1.5cm} p{1.5cm}}
		\toprule
		\multirow{4}{*}{\diagbox[height=4\line,innerwidth=5.5cm]{Submission body}{Model}} & \multicolumn{2}{c}{\makecell{Prop. scorer\\base}} & \multicolumn{2}{c}{\makecell{Prop. scorer\\gender neutralized} } \\
		\cmidrule(lr){2-3}
		\cmidrule(lr){4-5}
	& \makecell[c]{Original\\text}
        & \makecell[c]{Gender\\swap} 
        & \makecell[c]{Original\\text} 
        & \makecell[c]{Gender\\swap} \\ 
        \midrule
        {\tiny I am pregnant and live in a different country from my \underline{\textbf{mom}}. Due to the pandemic, we can't visit each other, and \underline{\textbf{she}} feels like \underline{\textbf{she}}'s missing out on my pregnancy. We've been in touch daily through calls, texts, and emails. \underline{\textbf{She}} wants to video chat to see my baby bump, which makes me uncomfortable as it triggers past anorexia-related anxieties. I've explained this to \underline{\textbf{her}}, but \underline{\textbf{she}} keeps asking. Given \underline{\textbf{her}} history of encouraging my past eating disorder, WIBTA if I set a firm boundary and refused to share baby bump pictures or video chats?}            & \makecell[c]{0.03}        & \makecell[c]{0.03}        & \makecell[c]{0.02}       & \makecell[c]{0.03}        \\
        \midrule
        {\tiny My dad recently passed away, possibly due to alcohol abuse. I had a strained relationship with him, unlike my younger sister W and older half-sister MJ, who were very close to him. He favored them and was disappointed in my life choices, eventually kicking me out when I was 18. I've since turned my life around, but now S wants me to attend his funeral. I'm conflicted because I don't want to go, but my \underline{\textbf{wife}} thinks I should support W and introduce our daughters to the family. WIBTA if I refused?}  & \makecell[c]{0.96}        & \makecell[c]{0.05}  & \makecell[c]{0.26}        & \makecell[c]{0.25}        \\
	\bottomrule
	\end{tabular}
	\label{tab:examples_gender_neutralization}
\end{table}

To mitigate the dependence of the model on these spurious correlations, we apply a gender neutralization strategy during the training of the propensity scorer~\cite{hall2019name}.
This strategy consists of identifying all the gendered words in a submission and swapping them with $50\%$ probability dynamically during training to neutraliza all male-female biases for gendered words.
We use the pairs of gendered words by~\citet{lu2020gender}.
When the gender neutralization strategy is applied to train a new model, the propensity of the second example discussed above does not change substantially under the replacement of the word ``wife'' (see the right part of \Cref{tab:examples_gender_neutralization}).
In addition, it returns a less extreme gender typicality for the submission that opens the possibility of finding a match for this story.

\begin{figure}
    \centering
    \includegraphics[width=1.\textwidth]{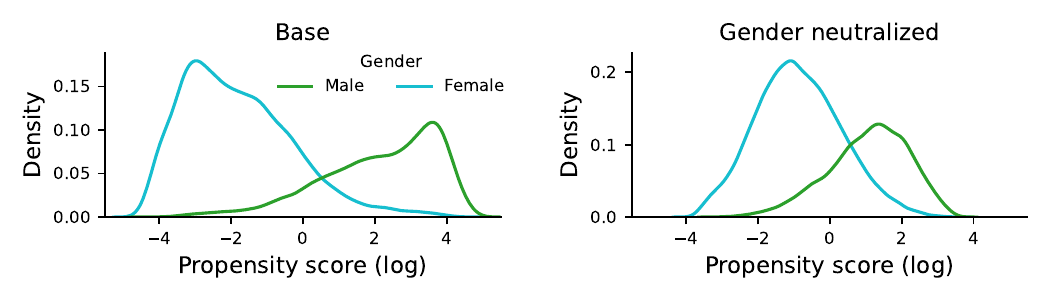}
    \caption{Distribution of the propensity score of male- and female-authored submissions obtained with the base (left) and gender neutralized propensity scorer (right). 
    }
    \label{fig:app_distribution_prop_scores}
\end{figure}

\begin{figure}
    \centering
    \includegraphics[width=0.99\textwidth]{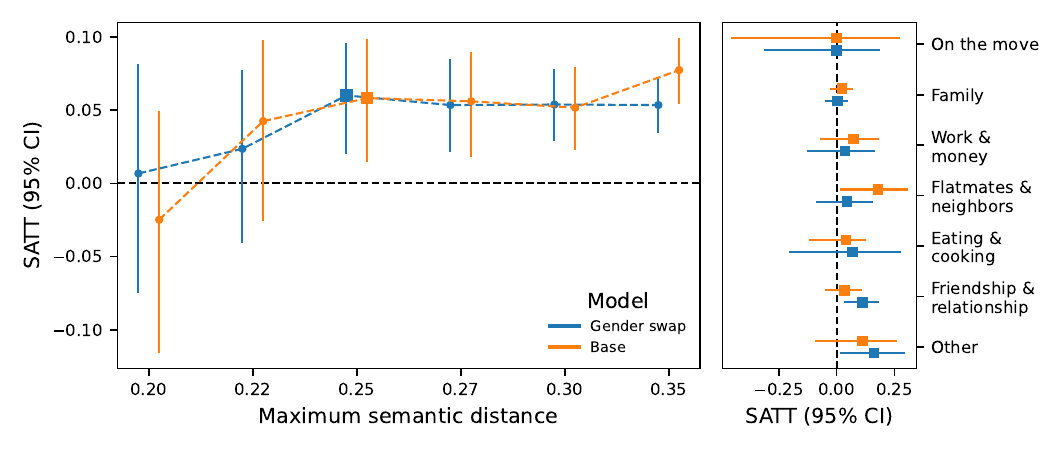}
    \caption{Comparison of the causal effect estimated with the gender neutralized and base propensity scorer models. 
    (a) Sample average treatment effect on the treated (SATT) for different values of the maximum semantic distance. 
    (b) SATT for each topic corresponding to the matches obtained with a maximum semantic distance equal to $0.25$.
    Vertical bars correspond to $95\%$ confidence intervals via bootstrap.}
    \label{fig:satt_both_prop_scorers}
\end{figure}

\Cref{fig:app_distribution_prop_scores} shows the distribution of propensity score for all submissions in the dataset separately by the gender of the author.
The gender neutralization strategy results in a desirable increase of overlap between the two distributions, suggesting that gendered words have a non-negligible correlation with the gender of authors in the \R{AITA} dataset.
\Cref{fig:satt_both_prop_scorers} shows the estimate of the causal effect obtained with both propensity score models. 
There are no major differences in the SATT between the two models.

\begin{figure}[]
    \centering
    \includegraphics[width=0.99\textwidth]{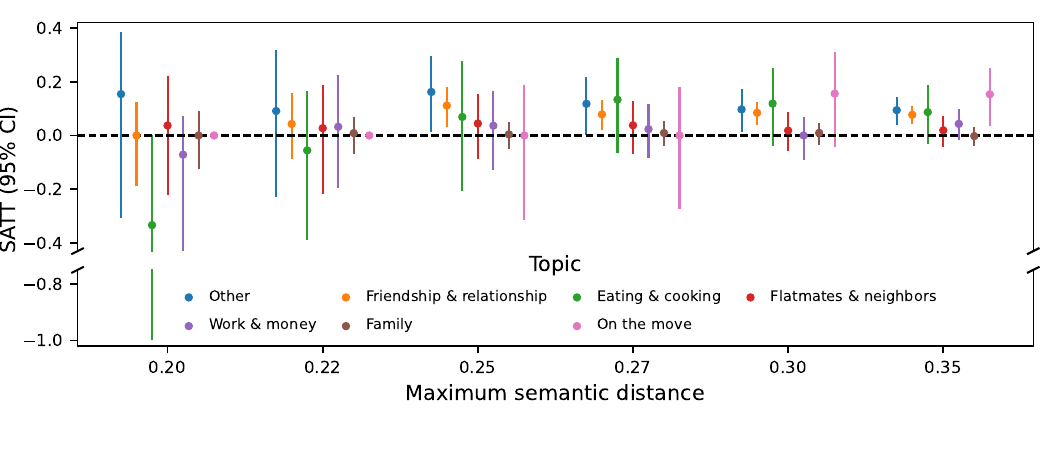}
    \caption{Average treatment effect on the treated (ATT) as a function of the maximum matching distance for each topic separately. Vertical bars correspond to bootstrap 95\% confidence intervals. 
    }
    \label{fig:att_topics}
\end{figure}

\section{Manual evaluation}

\subsection{Examples of matched submissions}

\clearpage

\begin{longtable}{|p{4.9cm}|p{4.9cm}|p{1.7cm}|}
	\caption{Examples of matched submissions. The third column reports the evaluations given by the annotators.} \label{tab:examples_matches}\\
     \hline \multicolumn{1}{|c|}{\textbf{Submission 1}} & \multicolumn{1}{c|}{\textbf{Submission 2}} & \multicolumn{1}{c|}{\textbf{Evaluations}} \\ \hline 
    \endfirsthead
    
    \multicolumn{3}{c}%
    {{\bfseries \tablename\ \thetable{} -- continued from previous page}} \\
    \hline \multicolumn{1}{|c|}{\textbf{Submission 1}} & \multicolumn{1}{c|}{\textbf{Submission 2}} & \multicolumn{1}{c|}{\textbf{Evaluations}} \\ \hline 
    \endhead
    
    \hline \multicolumn{3}{|r|}{{Continued on next page}} \\ \hline
    \endfoot
    
    \hline \hline
    \endlastfoot

  {\tiny Today, my mom [38F] asked me [17M] for 50PLN. It's a lot for me since I have limited pocket money. She's unemployed and looking for a part-time office job. She avoids telling me why she needs money, often for cigarettes, which I refuse to support. She already owes me 730PLN and hasn't repaid for over a year. When I offered to buy groceries instead, she stopped talking to me. She borrows from family and lenders but pays them back, making me feel unimportant. After confronting her, she accused me of thinking she's a failure. Am I wrong for refusing? How can I avoid worsening the conflict?} & {\tiny I (18F) work part-time at minimum wage and still live at home. My mom often borrows my car or money. I lent her \$30 last week, but she complains whenever I ask for it back. Today, she messaged me saying a family member can't fly in for Thanksgiving without \$100 for luggage and asked if I could spare it until next Friday. I told her no, and she called me a shitty person. Am I the asshole?} & \makecell{\small Very similar \\ \\\small Very similar \\ \\\small Very similar} \\
		 \hline
   {\tiny Before my best friend (17F) got a boyfriend, we hung out all the time. But after she started dating, she began ignoring me to spend time with him and his friends. When I talked to her about it, she accused me of being jealous and left college to be with him, leaving me alone. As we grew distant, we argued more. I started spending time with other friends, and she got upset, accusing me of excluding her. She blocked me on everything and convinced our friends to drop me. I sent her a message saying I was tired of her behavior, only for her to call me a "shit friend." Am I wrong for feeling this way?} & {\tiny I (17M) had a close friendship with a girl (18F) who recently started getting close to a guy I dislike. We've texted every day for years and could always talk out our problems. Now, she's phasing me and other friends out to spend time with him. I told her it upset me, but she ignored my messages. On a recent free day, she acted like she didn't know me. When I sent a "we need to talk" text, she left me on read. Frustrated, I messaged, "I'm just gonna assume you didn't have time to answer so I don't get really pissed off." She blocked me without replying. Mutual friends think I'm overreacting, but it feels like I always make sacrifices in our friendship. I don't think I deserved to be blocked for wanting to talk. Am I the asshole for being upset she's willing to throw away our years of friendship for a guy she hardly knows?} & \makecell{\small Very similar \\ \\\small Very similar \\ \\\small Very similar} \\
   \hline
   {\tiny Last week, I (16M) visited home and spent time with my girlfriend (16F) for the first time in two months. The last night I could see her was Saturday, but she chose to go to a party instead. She invited me, but I declined because I don't enjoy parties. I told her I wanted to spend that night together since it would be a month before we saw each other again. She declined, saying the party had been planned for weeks, which she only informed me of the night before. I was upset because of our limited time together, while she was upset that I expected her to devote the whole week to me. Am I the asshole for wanting to spend time with her before I leave despite her plans?} & {\tiny My girlfriend (19F) and I (19F) haven't seen much of each other in 3 months due to her externship. She got back a few days ago, but we've both been busy with work and family. Today, she was supposed to get off work at 9pm. I planned to see the new Star Wars movie with my family and be home by 10pm, hoping we'd spend some time together before bed. After the movie, I saw she went out with a mutual friend instead. I don't mind her hanging out with others, but now she won't be home for hours, and I'll likely be asleep when she gets back. Am I the asshole for being upset she went out instead of coming home?} & \makecell{\small Very similar \\ \\\small Somewhat \\ \small similar \\ \\\small Somewhat \\ \small similar} \\
   \hline
   {\tiny I (19F) was in a relationship with a girl for a year and a half. Things were mostly fine, but by June, I felt my romantic feelings for her had faded. I decided to be honest and told her I wanted to remain friends. She blocked me but later unblocked and sent messages saying she hated me and felt abandoned. She accused me of making her suicidal and not caring about her during her rough patch. Both of us have struggled with depression, and her messages blaming me were affecting my own mental health. I stepped away to protect myself from the constant negativity. Am I selfish or an asshole for stepping away even though I initially wanted to be there for her?} & {\tiny I've been struggling with severe depression and anxiety after failing my exams and not getting into university. My girlfriend (19F) broke up with me after I told her she was the only thing stopping me from seriously hurting myself. We're part of a small friendship group that supported our relationship. When she went to university, she kissed someone else and said she felt "trapped." I had a mental breakdown during a visit, and she later ended things, saying I needed to help myself first. I told her I need to distance myself for my mental health, but this means avoiding our entire friend group. Am I the asshole for wanting to cut them out of my life until I feel better, even if it seems selfish?} & \makecell{\small Neither \\ \small dissimilar \\ \small nor similar \\ \\ \small Somewhat \\ \small similar \\ \\ \small Somewhat \\ \small dissimilar \\ \\ \small Resolved as: \\ \small Neither \\ \small dissimilar \\ \small nor similar} \\
   \hline
   {\tiny I'm (20F) socially anxious and often use sarcasm to cope. A friend (23M) I've had an on-off friends-with-benefits relationship with recently told me my sarcastic attitude was giving him anxiety. I decided to be vulnerable and open about my feelings, despite feeling uncomfortable. Mid-conversation, he blocked me on that platform. I felt ashamed and embarrassed, like I was coerced into vulnerability only to be hurt. A few days ago, he said he wanted to know if he upset me, so I messaged him on our usual platform, asking what happened. This led to an argument where he felt I was blaming him for being uncomfortable. Am I the asshole for being upset and wanting to discuss my feelings, or am I being selfish and playing the victim?} & {\tiny Me (M22 at the time) and my best friend (M26 at the time) haven't spoken for 3 years because I cut him off. In 2017, we were inseparable. When my mother was diagnosed with cancer, I withdrew to spend time with her. He was supportive. After my mother passed, he made an offensive mom joke within 3 months. When I told him it was not cool, he dismissed it as "just something dudes do." I sought an apology, but he insisted it meant nothing. I cut him off and distanced myself from our mutual friends. Recently, he lost his job due to COVID, and I'm considering recommending him for a job. Am I the asshole for cutting him off and distancing myself from others, and would it be too much to offer him a job interview now?} & \makecell{\small Very \\ \small dissimilar \\ \\ \small Somewhat \\ \small dissimilar \\ \\ \small Somewhat \\ \small dissimilar} \\
   \hline
    {\tiny I'm 17M, and my stepsister is 18F. We've known each other for years but aren't close. She moved out recently, and our parents are going on vacation soon. They want her to come back and look after me. I don't mind her coming, but it's a 4-5 hour drive, and she just moved out. I feel bad making her do this just so our parents can force us to bond. I tried to convince my parents it's unnecessary, but they insist she comes so I won't be lonely. When I asked my stepsister, she said she doesn't want to but feels obligated because our parents pay for her phone and car. I feel guilty that she's being forced into this. AITA for trying to tell my parents that my stepsister looking after me is unnecessary?} & {\tiny I (19F) live with my boyfriend, best friend, and her girlfriend. We split the rent. Recently, my stepsister got accepted to a college nearby and wants to move in with us. My dad and his wife offered to pay her share of the rent, thinking it would help us bond, but I said no. My dad's wife called, saying it was wrong and cruel to turn her away and that her daughter wanted to bond as adults since we never had a sibling relationship growing up. I told her I didn't want to live with my stepsister and wanted to keep my current living situation. AITA for saying no without consideration? My dad and his wife think I'm being unfair, but I don't see her as family or a friend, so I don't feel bad.} & \makecell{\small Somewhat \\ \small dissimilar \\ \\ \small Somewhat \\ \small dissimilar \\ \\ \small Somewhat \\ \small dissimilar} \\
\end{longtable}

\subsection{Additional results of the manual evaluation}
\label{app:sec_agency_vs_gender_and_judgment}
The main purpose of the manual evaluation is to validate the matching procedure that identifies submissions describing similar situations.
In addition to this, we ask the annotators to evaluate the level of agency of the authors of the submissions, as described in the Methods section.
The analysis of the authors' agency in relation to their gender and moral judgment provides validation to the manual annotation and supports the discussion of our findings.

First, we test to what extent the level of agency is related to the self-disclosed gender of the author and the moral judgment received.
We fit two linear mixed-effects models:
$$ \text{Initiator\_score} \sim \text{Gender} * \text{isDissimilar} + \left(1 \vert \text{Annotator} \right) $$
$$ \text{Initiator\_score} \sim \text{Judgment} * \text{isDissimilar} + \left(1 \vert \text{Annotator} \right) $$
where $\text{isDissimilar}$ is a dummy variable that has value 1 if the submission belongs to a match evaluated as dissimilar, the term $\left(1 \vert \text{Annotator} \right)$ refers to a random intercept associated to each annotator, and the ``$*$'' indicates that the model considers both variables and their interaction.
The $\text{isDissimilar}$ variable indicates whether the submission belongs to the group of submissions that have a match considered to be similar.
The coefficient corresponding to the interaction term estimates the differential effect of the predictor (i.e., gender and judgment) in the subset of submissions that result from a wrong match.

We present the result of the first model in \Cref{app:regress_initiator_on_gender}.
Although the model returns no significant overall contribution of the gender (``Gender''), the interaction term between gender and isDissimilar has a positive and statistically significant coefficient (``Gender:isDissimilar''), meaning that male authors have a higher initiator score when the submission belongs to a match judged to be dissimilar.
In plain terms, there is no significant difference in the level of agency of male and female authors for submissions judged to be similar, whereas male authors have a higher level of agency for dissimilar submissions.
This result suggests that the level of agency of the protagonist is effectively controlled in the subset of correctly matched submissions, whereas male protagonists are more agentic than female ones in the subset of submissions with a dissimilar match.

\begin{table}
\caption{OLS regression model of author's gender on initiator score.}
\label{app:regress_initiator_on_gender}
\begin{center}
\begin{tabular}{p{3cm}p{2cm}p{3.5cm}p{2.7cm}}
\hline
Model:            & MixedLM & Dependent Variable: & Initiator\_score  \\
No. Observations: & 600     & Method:             & REML                  \\
No. Groups:       & 5       & Scale:              & 1.1004                \\
Min. group size:  & 120     & Log-Likelihood:     & -889.0219             \\
Max. group size:  & 120     & Converged:          & Yes                   \\
Mean group size:  & 120.0   &                     &                       \\
\hline
\end{tabular}
\end{center}

\begin{center}
\begin{tabular}{p{3cm}p{1.2cm}p{1.2cm}p{1.2cm}p{1.2cm}p{1.2cm}p{1.2cm}}
\hline
          &  \makecell[c]{Coef.} & \makecell[c]{Std.Err.} &      \makecell[c]{z} & \makecell[c]{P$> |$z$|$} & [0.025 & 0.975]  \\
\hline
Intercept &  1.567 &    0.159 &  9.884 &       0.000 &  1.256 &  1.877  \\
Gender    &  0.095 &    0.108 &  0.883 &       0.377 & -0.116 &  0.307  \\
isDissimilar        & -0.262 &    0.126 & -2.078 &       0.038 & -0.508 & -0.015  \\
Gender:isDissimilar &  0.427 &    0.177 &  2.409 &       0.016 &  0.080 &  0.775  \\
Group Var &  0.096 &    0.071 &        &             &        &         \\
\hline
\end{tabular}
\end{center}
\end{table}

\Cref{app:regress_initiator_on_judgment} reports the result of the second model.
Here, negative moral judgment is associated to a higher level of agency (``Judgment''), while there is no significant differential effect with respect to the fact that the submission belongs to a match judged dissimilar (``Judgment:isDissimilar'').
This result points to higher levels of agency associated to a higher likelihood of receiving a negative moral judgment.

\begin{table}
\caption{OLS regression model of moral judgment on initiator score.}
\label{app:regress_initiator_on_judgment}
\begin{center}
\begin{tabular}{p{3cm}p{2cm}p{3.5cm}p{2.7cm}}
\hline
Model:            & MixedLM & Dependent Variable: & Initiator\_score  \\
No. Observations: & 600     & Method:             & REML                  \\
No. Groups:       & 5       & Scale:              & 1.0612                \\
Min. group size:  & 120     & Log-Likelihood:     & -877.7939             \\
Max. group size:  & 120     & Converged:          & Yes                   \\
Mean group size:  & 120.0   &                     &                       \\
\hline
\end{tabular}
\end{center}

\begin{center}
\begin{tabular}{p{3cm}p{1.2cm}p{1.2cm}p{1.2cm}p{1.2cm}p{1.2cm}p{1.2cm}}
\hline
          &  \makecell[c]{Coef.} & \makecell[c]{Std.Err.} &      \makecell[c]{z} & \makecell[c]{P$> |$z$|$} & [0.025 & 0.975]  \\
\hline
Intercept   &  1.482 &    0.152 &  9.741 &       0.000 &  1.184 &  1.781  \\
Judgment    &  0.692 &    0.135 &  5.128 &       0.000 &  0.428 &  0.957  \\
isDissimilar          & -0.019 &    0.097 & -0.201 &       0.841 & -0.209 &  0.170  \\
Judgment:isDissimilar & -0.101 &    0.227 & -0.444 &       0.657 & -0.545 &  0.344  \\
Group Var   &  0.098 &    0.074 &        &             &        &         \\
\hline
\end{tabular}
\end{center}
\end{table}

%
%
%
\bibliographystyle{splncs04nat}
\bibliography{biblio}

\begin{thebibliography}{73}
\providecommand{\natexlab}[1]{#1}
\providecommand{\url}[1]{\texttt{#1}}
\providecommand{\urlprefix}{URL }
\expandafter\ifx\csname urlstyle\endcsname\relax
  \providecommand{\doi}[1]{doi:\discretionary{}{}{}#1}\else
  \providecommand{\doi}{doi:\discretionary{}{}{}\begingroup
  \urlstyle{rm}\Url}\fi

\bibitem[{Ammari et~al.(2019)Ammari, Schoenebeck, and Romero}]{ammari2019self}
Ammari, T., Schoenebeck, S., Romero, D.: Self-declared throwaway accounts on
  reddit: How platform affordances and shared norms enable parenting disclosure
  and support. Proceedings of the ACM on Human-Computer Interaction
  \textbf{3}(CSCW), 1--30 (2019)

\bibitem[{Arnestad et~al.(2020)Arnestad, Studzinska, Nordmo, and
  Matthiesen}]{arnestad2020hetoo}
Arnestad, M., Studzinska, A., Nordmo, M., Matthiesen, S.B.: \#hetoo? men
  trivialize cases of sexual harassment by a female aggressor toward a male
  victim, but women do not (Aug 2020), \doi{10.31234/osf.io/q2zhg},
  \urlprefix\url{https://psyarxiv.com/q2zhg}

\bibitem[{Balsamo et~al.(2023)Balsamo, Bajardi, Morales, Monti, and
  Schifanella}]{balsamo2023pursuit}
Balsamo, D., Bajardi, P., Morales, G.D.F., Monti, C., Schifanella, R.: The
  pursuit of peer support for opioid use recovery on reddit. In: Proceedings of
  the International AAAI Conference on Web and Social Media, vol.~17, pp.
  12--23 (2023)

\bibitem[{Barbee et~al.(1993)Barbee, Cunningham, Winstead, Derlega, Gulley,
  Yankeelov, and Druen}]{barbee1993effects}
Barbee, A.P., Cunningham, M.R., Winstead, B.A., Derlega, V.J., Gulley, M.R.,
  Yankeelov, P.A., Druen, P.B.: Effects of {Gender} {Role} {Expectations} on
  the {Social} {Support} {Process}. Journal of Social Issues \textbf{49}(3),
  175--190 (Oct 1993), ISSN 0022-4537, 1540-4560,
  \doi{10.1111/j.1540-4560.1993.tb01175.x},
  \urlprefix\url{https://spssi.onlinelibrary.wiley.com/doi/10.1111/j.1540-4560.1993.tb01175.x}

\bibitem[{Barber and Odean(2001)}]{barber2001boys}
Barber, B.M., Odean, T.: Boys will be {Boys}: {Gender}, {Overconfidence}, and
  {Common} {Stock} {Investment}*. The Quarterly Journal of Economics
  \textbf{116}(1), 261--292 (Feb 2001), ISSN 0033-5533,
  \doi{10.1162/003355301556400},
  \urlprefix\url{https://doi.org/10.1162/003355301556400}

\bibitem[{Baronchelli(2024)}]{baronchelli2024shaping}
Baronchelli, A.: Shaping new norms for ai. Philosophical Transactions of the
  Royal Society B \textbf{379}(1897), 20230028 (2024)

\bibitem[{Baumgartner et~al.(2020)Baumgartner, Zannettou, Keegan, Squire, and
  Blackburn}]{Baumgartner_Zannettou_Keegan_Squire_Blackburn_2020}
Baumgartner, J., Zannettou, S., Keegan, B., Squire, M., Blackburn, J.: The
  pushshift reddit dataset. Proceedings of the International AAAI Conference on
  Web and Social Media \textbf{14}(1), 830--839 (May 2020),
  \urlprefix\url{https://ojs.aaai.org/index.php/ICWSM/article/view/7347}

\bibitem[{Blei et~al.(2001)Blei, Ng, and Jordan}]{blei2001latent}
Blei, D., Ng, A., Jordan, M.: Latent dirichlet allocation. In: Dietterich, T.,
  Becker, S., Ghahramani, Z. (eds.) Advances in Neural Information Processing
  Systems, vol.~14, MIT Press (2001),
  \urlprefix\url{https://proceedings.neurips.cc/paper/2001/file/296472c9542ad4d4788d543508116cbc-Paper.pdf}

\bibitem[{Botzer et~al.(2022)Botzer, Gu, and Weninger}]{botzer2022analysis}
Botzer, N., Gu, S., Weninger, T.: Analysis of moral judgment on reddit. IEEE
  Transactions on Computational Social Systems pp. 1--11 (2022),
  \doi{10.1109/TCSS.2022.3160677}

\bibitem[{Byrnes et~al.(1999)Byrnes, Miller, and Schafer}]{byrnes1999gender}
Byrnes, J.P., Miller, D.C., Schafer, W.D.: Gender differences in risk taking:
  {A} meta-analysis. Psychological Bulletin \textbf{125}(3), 367--383 (1999),
  ISSN 1939-1455, \doi{10.1037/0033-2909.125.3.367}, place: US Publisher:
  American Psychological Association

\bibitem[{Chandrasekharan et~al.(2018)Chandrasekharan, Samory, Jhaver, Charvat,
  Bruckman, Lampe, Eisenstein, and Gilbert}]{chandrasekharan2018internet}
Chandrasekharan, E., Samory, M., Jhaver, S., Charvat, H., Bruckman, A., Lampe,
  C., Eisenstein, J., Gilbert, E.: The internet's hidden rules: An empirical
  study of reddit norm violations at micro, meso, and macro scales. Proc. ACM
  Hum.-Comput. Interact. \textbf{2}(CSCW) (nov 2018), \doi{10.1145/3274301},
  \urlprefix\url{https://doi.org/10.1145/3274301}

\bibitem[{Christensen and Gomila(2012)}]{christensen2012moral}
Christensen, J., Gomila, A.: Moral dilemmas in cognitive neuroscience of moral
  decision-making: A principled review. Neuroscience \& Biobehavioral Reviews
  \textbf{36}(4), 1249--1264 (2012), ISSN 0149-7634,
  \doi{https://doi.org/10.1016/j.neubiorev.2012.02.008},
  \urlprefix\url{https://www.sciencedirect.com/science/article/pii/S0149763412000346}

\bibitem[{Chu and Gr{\"u}hn(2018)}]{chu2018moral}
Chu, Q., Gr{\"u}hn, D.: Moral {Judgments} and {Social} {Stereotypes}: {Do} the
  {Age} and {Gender} of the {Perpetrator} and the {Victim} {Matter}? Social
  Psychological and Personality Science \textbf{9}(4), 426--434 (May 2018),
  ISSN 1948-5506, 1948-5514, \doi{10.1177/1948550617711226},
  \urlprefix\url{http://journals.sagepub.com/doi/10.1177/1948550617711226}

\bibitem[{Cinelli et~al.(2021)Cinelli, De~Francisci~Morales, Galeazzi,
  Quattrociocchi, and Starnini}]{cinelli2021echo}
Cinelli, M., De~Francisci~Morales, G., Galeazzi, A., Quattrociocchi, W.,
  Starnini, M.: The {Echo} {Chamber} {Effect} on {Social} {Media}. Proceedings
  of the National Academy of Sciences \textbf{118}(9), e2023301118 (2021),
  \doi{10.1073/pnas.2023301118},
  \urlprefix\url{http://www.pnas.org/lookup/doi/10.1073/pnas.2023301118}

\bibitem[{Coleman(1994)}]{coleman1994foundations}
Coleman, J.S.: Foundations of Social Theory. Harvard University Press (1994)

\bibitem[{Croson and Gneezy(2009)}]{croson2009gender}
Croson, R., Gneezy, U.: Gender differences in preferences. Journal of Economic
  literature \textbf{47}(2), 448--474 (2009),
  \urlprefix\url{https://www.aeaweb.org/articles?id=10.1257/jel.47.2.448},
  publisher: American Economic Association

\bibitem[{De~Candia et~al.(2022)De~Candia, De~Francisci~Morales, Monti, and
  Bonchi}]{Candia2022Social}
De~Candia, S., De~Francisci~Morales, G., Monti, C., Bonchi, F.: Social norms on
  reddit: A demographic analysis. In: 14th ACM Web Science Conference 2022, pp.
  139--147, WebSci '22, Association for Computing Machinery, New York, NY, USA
  (2022), ISBN 9781450391917, \doi{10.1145/3501247.3531549},
  \urlprefix\url{https://doi.org/10.1145/3501247.3531549}

\bibitem[{De~Choudhury and De(2014)}]{de2014mental}
De~Choudhury, M., De, S.: Mental health discourse on reddit: Self-disclosure,
  social support, and anonymity. In: Proceedings of the international AAAI
  conference on web and social media, vol.~8, pp. 71--80 (2014)

\bibitem[{Devlin et~al.(2018)Devlin, Chang, Lee, and
  Toutanova}]{devlin2018bert}
Devlin, J., Chang, M.W., Lee, K., Toutanova, K.: Bert: Pre-training of deep
  bidirectional transformers for language understanding (2018),
  \doi{10.48550/ARXIV.1810.04805},
  \urlprefix\url{https://arxiv.org/abs/1810.04805}

\bibitem[{Durkheim(1895)}]{durkheim1985regles}
Durkheim, {\'E}.: Les r{\`e}gles de la m{\'e}thode sociologique. Flammarion
  (1895)

\bibitem[{Eagly(1995)}]{eagly1995science}
Eagly, A.H.: The science and politics of comparing women and men. American
  Psychologist \textbf{50}(3), 145--158 (Mar 1995), ISSN 1935-990X, 0003-066X,
  \doi{10.1037/0003-066X.50.3.145},
  \urlprefix\url{http://doi.apa.org/getdoi.cfm?doi=10.1037/0003-066X.50.3.145}

\bibitem[{Eisenchlas(2013)}]{eisenchlas2013gender}
Eisenchlas, S.A.: Gender roles and expectations: Any changes online? SAGE Open
  \textbf{3}(4), 2158244013506446 (2013), \doi{10.1177/2158244013506446},
  \urlprefix\url{https://doi.org/10.1177/2158244013506446}

\bibitem[{Elinder and Erixson(2012)}]{elinder2012gender}
Elinder, M., Erixson, O.: Gender, social norms, and survival in maritime
  disasters. Proceedings of the National Academy of Sciences \textbf{109}(33),
  13220--13224 (2012), \doi{10.1073/pnas.1207156109},
  \urlprefix\url{https://www.pnas.org/doi/abs/10.1073/pnas.1207156109}

\bibitem[{Fitzgerald(2016)}]{fitzgerald2016social}
Fitzgerald, K.J. (ed.): Social roles and social norms. Social issues, justice
  and status, Nova Publishers, New York (2016), ISBN 978-1-63483-952-5

\bibitem[{Giorgi et~al.(2023)Giorgi, Zhao, Feng, and Martin}]{giorgi2023author}
Giorgi, S., Zhao, K., Feng, A.H., Martin, L.J.: Author as {Character} and
  {Narrator}: {Deconstructing} {Personal} {Narratives} from the
  r/{AmITheAsshole} {Reddit} {Community}. Proceedings of the International AAAI
  Conference on Web and Social Media \textbf{17}, 233--244 (Jun 2023), ISSN
  2334-0770, 2162-3449, \doi{10.1609/icwsm.v17i1.22141},
  \urlprefix\url{https://ojs.aaai.org/index.php/ICWSM/article/view/22141}

\bibitem[{Gray and Wegner(2009)}]{gray2009moral}
Gray, K., Wegner, D.M.: Moral typecasting: Divergent perceptions of moral
  agents and moral patients. Journal of Personality and Social Psychology
  \textbf{96}(3), 505--520 (2009), \doi{10.1037/a0013748}

\bibitem[{Greifer and Stuart(2023)}]{greifer2023choosing}
Greifer, N., Stuart, E.A.: Choosing the causal estimand for propensity score
  analysis of observational studies (2023)

\bibitem[{Greve-Poulsen et~al.(2023)Greve-Poulsen, Larsen, Pedersen, and
  Alb{\ae}k}]{greve-poulsen2023no}
Greve-Poulsen, K., Larsen, F.K., Pedersen, R.T., Alb{\ae}k, E.: No {Gender}
  {Bias} in {Audience} {Perceptions} of {Male} and {Female} {Experts} in the
  {News}: {Equally} {Competent} and {Persuasive}. The International Journal of
  Press/Politics \textbf{28}(1), 116--137 (Jan 2023), ISSN 1940-1612,
  \doi{10.1177/19401612211025499},
  \urlprefix\url{https://doi.org/10.1177/19401612211025499}, publisher: SAGE
  Publications Inc

\bibitem[{Ho et~al.(2007)Ho, Imai, King, and Stuart}]{ho2007matching}
Ho, D.E., Imai, K., King, G., Stuart, E.A.: Matching as nonparametric
  preprocessing for reducing model dependence in parametric causal inference.
  Political Analysis \textbf{15}(3), 199--236 (2007), \doi{10.1093/pan/mpl013}

\bibitem[{Imai et~al.(2008)Imai, King, and Stuart}]{imai2008misunderstandings}
Imai, K., King, G., Stuart, E.A.: Misunderstandings between experimentalists
  and observationalists about causal inference. Journal of the Royal
  Statistical Society: Series A (Statistics in Society) \textbf{171}(2),
  481--502 (2008), \doi{https://doi.org/10.1111/j.1467-985X.2007.00527.x},
  \urlprefix\url{https://rss.onlinelibrary.wiley.com/doi/abs/10.1111/j.1467-985X.2007.00527.x}

\bibitem[{Imbens(2004)}]{imbens2004nonparametric}
Imbens, G.W.: {Nonparametric Estimation of Average Treatment Effects Under
  Exogeneity: A Review}. The Review of Economics and Statistics \textbf{86}(1),
  4--29 (02 2004), ISSN 0034-6535, \doi{10.1162/003465304323023651},
  \urlprefix\url{https://doi.org/10.1162/003465304323023651}

\bibitem[{Johnson et~al.(2006)Johnson, McDermott, Barrett, Cowden, Wrangham,
  McIntyre, and Peter~Rosen}]{johnson2006overconfidence}
Johnson, D.D., McDermott, R., Barrett, E.S., Cowden, J., Wrangham, R.,
  McIntyre, M.H., Peter~Rosen, S.: Overconfidence in wargames: experimental
  evidence on expectations, aggression, gender and testosterone. Proceedings of
  the Royal Society B: Biological Sciences \textbf{273}(1600), 2513--2520
  (2006), \doi{10.1098/rspb.2006.3606},
  \urlprefix\url{https://royalsocietypublishing.org/doi/abs/10.1098/rspb.2006.3606}

\bibitem[{Keith et~al.(2020)Keith, Jensen, and O{'}Connor}]{keith2020text}
Keith, K., Jensen, D., O{'}Connor, B.: Text and causal inference: A review of
  using text to remove confounding from causal estimates. In: Jurafsky, D.,
  Chai, J., Schluter, N., Tetreault, J. (eds.) Proceedings of the 58th Annual
  Meeting of the Association for Computational Linguistics, pp. 5332--5344,
  Association for Computational Linguistics, Online (Jul 2020),
  \doi{10.18653/v1/2020.acl-main.474},
  \urlprefix\url{https://aclanthology.org/2020.acl-main.474}

\bibitem[{Klemm et~al.(1999)Klemm, Hurst, Dearholt, and
  Trone}]{klemm1999gender}
Klemm, P., Hurst, M., Dearholt, S.L., Trone, S.R.: Gender differences on
  {Internet} cancer support groups. Computers in Nursing \textbf{17}(2), 65--72
  (1999), ISSN 0736-8593

\bibitem[{Kneavel(2021)}]{kneavel2021relationship}
Kneavel, M.: Relationship {Between} {Gender}, {Stress}, and {Quality} of
  {Social} {Support}. Psychological Reports \textbf{124}(4), 1481--1501 (Aug
  2021), ISSN 0033-2941, 1558-691X, \doi{10.1177/0033294120939844},
  \urlprefix\url{http://journals.sagepub.com/doi/10.1177/0033294120939844}

\bibitem[{Kreager et~al.(2016)Kreager, Staff, Gauthier, Lefkowitz, and
  Feinberg}]{kreager2016double}
Kreager, D.A., Staff, J., Gauthier, R., Lefkowitz, E.S., Feinberg, M.E.: The
  double standard at sexual debut: Gender, sexual behavior and adolescent peer
  acceptance. Sex Roles \textbf{75}(7), 377--392 (Oct 2016), ISSN 1573-2762,
  \doi{10.1007/s11199-016-0618-x},
  \urlprefix\url{https://doi.org/10.1007/s11199-016-0618-x}

\bibitem[{Leiby et~al.(2021)Leiby, Bos, and Krain}]{leiby2021gendered}
Leiby, M., Bos, A.L., Krain, M.: Gendered framing in human rights campaigns.
  Journal of Human Rights \textbf{20}(3), 263--281 (2021),
  \doi{10.1080/14754835.2021.1882837},
  \urlprefix\url{https://doi.org/10.1080/14754835.2021.1882837}

\bibitem[{Li et~al.(2019)Li, Coduto, and Morr}]{li2019communicating}
Li, S., Coduto, K.D., Morr, L.: Communicating social support online: The roles
  of emotional disclosures and gender cues in support provision. Telematics and
  Informatics \textbf{39}, 92--100 (2019)

\bibitem[{Lindsey(2015)}]{lindsey2015gender}
Lindsey, L.: Gender roles: {A} sociological perspective. Taylor \& Francis
  (2015), ISBN 978-1-317-34808-5

\bibitem[{Low et~al.(2023)Low, Lavin, Du, and Fang}]{low2023riskinformed}
Low, B., Lavin, D., Du, C.R., Fang, C.: Risk-{Informed} and {AI}-{Based} {Bias}
  {Detection} on {Gender}, {Race}, and {Income} {Using} {Gen}-{Z} {Survey}
  {Data}. IEEE Access \textbf{11}, 88317--88328 (2023), ISSN 2169-3536,
  \doi{10.1109/ACCESS.2023.3305636},
  \urlprefix\url{https://ieeexplore.ieee.org/document/10220079/}

\bibitem[{Lu et~al.(2020)Lu, Mardziel, Wu, Amancharla, and
  Datta}]{lu2020gender}
Lu, K., Mardziel, P., Wu, F., Amancharla, P., Datta, A.: Gender Bias in Neural
  Natural Language Processing, pp. 189--202. Springer International Publishing,
  Cham (2020), ISBN 978-3-030-62077-6, \doi{10.1007/978-3-030-62077-6_14},
  \urlprefix\url{https://doi.org/10.1007/978-3-030-62077-6_14}

\bibitem[{Mah et~al.(2014)Mah, Taylor, Hoang, and Cook}]{mah2014using}
Mah, C.L., Taylor, E., Hoang, S., Cook, B.: Using vignettes to tap into moral
  reasoning in public health policy: practical advice and design principles
  from a study on food advertising to children. Am. J. Public Health
  \textbf{104}(10), 1826--1832 (Oct 2014)

\bibitem[{Maudslay et~al.(2019)Maudslay, Gonen, Cotterell, and
  Teufel}]{hall2019name}
Maudslay, R.H., Gonen, H., Cotterell, R., Teufel, S.: It{'}s all in the name:
  Mitigating gender bias with name-based counterfactual data substitution. In:
  Inui, K., Jiang, J., Ng, V., Wan, X. (eds.) Proceedings of the 2019
  Conference on Empirical Methods in Natural Language Processing and the 9th
  International Joint Conference on Natural Language Processing (EMNLP-IJCNLP),
  pp. 5267--5275, Association for Computational Linguistics, Hong Kong, China
  (Nov 2019), \doi{10.18653/v1/D19-1530},
  \urlprefix\url{https://aclanthology.org/D19-1530}

\bibitem[{Miller(2020)}]{miller2020investigating}
Miller, B.: Investigating reddit self-disclosure and confessions in relation to
  connectedness, social support, and life satisfaction. Social media and
  society \textbf{9}, 39--62 (2020)

\bibitem[{Mok et~al.(2023)Mok, Inzlicht, and Anderson}]{mok2023echo}
Mok, L., Inzlicht, M., Anderson, A.: Echo {Tunnels}: {Polarized} {News}
  {Sharing} {Online} {Runs} {Narrow} but {Deep}. Proceedings of the
  International AAAI Conference on Web and Social Media \textbf{17}, 662--673
  (Jun 2023), ISSN 2334-0770, \doi{10.1609/icwsm.v17i1.22177},
  \urlprefix\url{https://ojs.aaai.org/index.php/ICWSM/article/view/22177}

\bibitem[{Mozer et~al.(2020)Mozer, Miratrix, Kaufman, and
  Jason~Anastasopoulos}]{mozer2020matching}
Mozer, R., Miratrix, L., Kaufman, A.R., Jason~Anastasopoulos, L.: Matching with
  text data: An experimental evaluation of methods for matching documents and
  of measuring match quality. Political Analysis \textbf{28}(4), 445--468
  (2020), \doi{10.1017/pan.2020.1}

\bibitem[{Mulac et~al.(2006)Mulac, Bradac, and Gibbons}]{mulac2006empirical}
Mulac, A., Bradac, J., Gibbons, P.: Empirical support for the gender-as-culture
  hypothesis.: {An} intercultural analysis of male/female language differences.
  Human Communication Research \textbf{27}(1), 121--152 (Jan 2006), ISSN
  03603989, \doi{10.1111/j.1468-2958.2001.tb00778.x},
  \urlprefix\url{https://academic.oup.com/hcr/article/27/1/121-152/4554820}

\bibitem[{Muthusamy et~al.(2023)Muthusamy, Rizk, Kate, Venkateswaran,
  Isahagian, Gulati, and Dube}]{muthusamy2023towards}
Muthusamy, V., Rizk, Y., Kate, K., Venkateswaran, P., Isahagian, V., Gulati,
  A., Dube, P.: Towards large language model-based personal agents in the
  enterprise: Current trends and open problems. In: The 2023 Conference on
  Empirical Methods in Natural Language Processing (2023)

\bibitem[{Nguyen et~al.(2022)Nguyen, Lyall, Tran, Shin, Carroll, Klein, and
  Xie}]{nguyen2022mapping}
Nguyen, T.D., Lyall, G., Tran, A., Shin, M., Carroll, N.G., Klein, C., Xie, L.:
  Mapping topics in 100,000 real-life moral dilemmas. Proceedings of the
  International AAAI Conference on Web and Social Media \textbf{16}(1),
  699--710 (May 2022),
  \urlprefix\url{https://ojs.aaai.org/index.php/ICWSM/article/view/19327}

\bibitem[{Parrigon et~al.(2017)Parrigon, Woo, Tay, and
  Wang}]{parrigon2017captioning}
Parrigon, S., Woo, S.E., Tay, L., Wang, T.: {CAPTION}-ing the situation: A
  lexically-derived taxonomy of psychological situation characteristics.
  Journal of Personality and Social Psychology \textbf{112}(4), 642--681 (apr
  2017), \doi{10.1037/pspp0000111}

\bibitem[{Parsons(1937)}]{parsons1937structure}
Parsons, T.: The Structure of Social Action. Free Press New York (1937)

\bibitem[{Patil et~al.(2024)Patil, Dhotre, Gawande, Mate, Shelke, and
  Bhoye}]{patil2024transformative}
Patil, D.D., Dhotre, D.R., Gawande, G.S., Mate, D.S., Shelke, M.V., Bhoye,
  T.S.: Transformative trends in generative ai: Harnessing large language
  models for natural language understanding and generation. International
  Journal of Intelligent Systems and Applications in Engineering
  \textbf{12}(4s), 309--319 (2024)

\bibitem[{Pino et~al.(2020)Pino, Zhang, and Wang}]{pino2020hearty}
Pino, G., Zhang, C.X., Wang, Z.: ``(s)he's so hearty'': Gender cues,
  stereotypes, and expectations of warmth in peer-to-peer accommodation
  services. International Journal of Hospitality Management \textbf{91}, 102650
  (2020), ISSN 0278-4319, \doi{https://doi.org/10.1016/j.ijhm.2020.102650},
  \urlprefix\url{https://www.sciencedirect.com/science/article/pii/S0278431920302024}

\bibitem[{Rauthmann et~al.(2014)Rauthmann, Gallardo-Pujol, Guillaume, Todd,
  Nave, Sherman, Ziegler, Jones, and Funder}]{rauthmann2014situational}
Rauthmann, J.F., Gallardo-Pujol, D., Guillaume, E., Todd, E., Nave, C.S.,
  Sherman, R.A., Ziegler, M., Jones, A.B., Funder, D.C.: The situational eight
  diamonds: a taxonomy of major dimensions of situation characteristics.
  Journal of personality and social psychology \textbf{107 4}, 677--718 (2014)

\bibitem[{{\v R}eh{\r u}{\v r}ek and Sojka(2010)}]{rehurek_lrec}
{\v R}eh{\r u}{\v r}ek, R., Sojka, P.: {Software Framework for Topic Modelling
  with Large Corpora}. In: {Proceedings of the LREC 2010 Workshop on New
  Challenges for NLP Frameworks}, pp. 45--50, ELRA, Valletta, Malta (May 2010),
  \url{http://is.muni.cz/publication/884893/en}

\bibitem[{Reimers and Gurevych(2019)}]{reimers2019sentence}
Reimers, N., Gurevych, I.: Sentence-bert: Sentence embeddings using siamese
  bert-networks. In: Proceedings of the 2019 Conference on Empirical Methods in
  Natural Language Processing, Association for Computational Linguistics (11
  2019), \urlprefix\url{https://arxiv.org/abs/1908.10084}

\bibitem[{Reynolds et~al.(2020)Reynolds, Howard, Sj{\aa}stad, Zhu, Okimoto,
  Baumeister, Aquino, and Kim}]{reynolds2020man}
Reynolds, T., Howard, C., Sj{\aa}stad, H., Zhu, L., Okimoto, T.G., Baumeister,
  R.F., Aquino, K., Kim, J.: Man up and take it: Gender bias in moral
  typecasting. Organizational Behavior and Human Decision Processes
  \textbf{161}, 120--141 (2020), ISSN 0749-5978,
  \doi{https://doi.org/10.1016/j.obhdp.2020.05.002},
  \urlprefix\url{https://www.sciencedirect.com/science/article/pii/S0749597820303630}

\bibitem[{ROSENBAUM and RUBIN(1983)}]{rosenbaum1983central}
ROSENBAUM, P.R., RUBIN, D.B.: {The central role of the propensity score in
  observational studies for causal effects}. Biometrika \textbf{70}(1), 41--55
  (04 1983), ISSN 0006-3444, \doi{10.1093/biomet/70.1.41},
  \urlprefix\url{https://doi.org/10.1093/biomet/70.1.41}

\bibitem[{Scott(1971)}]{scott1971internalization}
Scott, J.F.: Internalization of norms: A sociological theory of moral
  commitment. Prentice-Hall (1971)

\bibitem[{Stade et~al.(2024)Stade, Stirman, Ungar, Boland, Schwartz, Yaden,
  Sedoc, DeRubeis, Willer, and Eichstaedt}]{stade2024large}
Stade, E.C., Stirman, S.W., Ungar, L.H., Boland, C.L., Schwartz, H.A., Yaden,
  D.B., Sedoc, J., DeRubeis, R.J., Willer, R., Eichstaedt, J.C.: Large language
  models could change the future of behavioral healthcare: a proposal for
  responsible development and evaluation. npj Mental Health Research
  \textbf{3}(1), 12 (2024)

\bibitem[{Stuart(2010)}]{stuart2010matching}
Stuart, E.A.: Matching methods for causal inference: A review and a look
  forward. Statistical science : a review journal of the Institute of
  Mathematical Statistics \textbf{25}(1), 1--21 (Feb 2010), ISSN 0883-4237,
  \doi{10.1214/09-STS313}, \urlprefix\url{https://doi.org/10.1214/09-STS313},
  20871802[pmid]

\bibitem[{Sunstein(1996)}]{sunstein1996social}
Sunstein, C.R.: Social {Norms} and {Social} {Roles}. Columbia Law Review
  \textbf{96}(4), 903 (May 1996), ISSN 00101958, \doi{10.2307/1123430},
  \urlprefix\url{https://www.jstor.org/stable/1123430?origin=crossref}

\bibitem[{Tajfel(1973)}]{tajfel1973roots}
Tajfel, H.: The roots of prejudice: Cognitive aspects. Psychology and race pp.
  76--95 (1973)

\bibitem[{Tamres et~al.(2002)Tamres, Janicki, and Helgeson}]{tamres2002sex}
Tamres, L.K., Janicki, D., Helgeson, V.S.: Sex {Differences} in {Coping}
  {Behavior}: {A} {Meta}-{Analytic} {Review} and an {Examination} of {Relative}
  {Coping}. Personality and Social Psychology Review \textbf{6}(1), 2--30 (Feb
  2002), ISSN 1088-8683, 1532-7957, \doi{10.1207/S15327957PSPR0601_1},
  \urlprefix\url{http://journals.sagepub.com/doi/10.1207/S15327957PSPR0601_1}

\bibitem[{Tifferet(2020)}]{tifferet2020gender}
Tifferet, S.: Gender {Differences} in {Social} {Support} on {Social} {Network}
  {Sites}: {A} {Meta}-{Analysis}. Cyberpsychology, Behavior, and Social
  Networking \textbf{23}(4), 199--209 (Apr 2020), ISSN 2152-2715, 2152-2723,
  \doi{10.1089/cyber.2019.0516},
  \urlprefix\url{https://www.liebertpub.com/doi/10.1089/cyber.2019.0516}

\bibitem[{Tomita et~al.(2021)Tomita, Barense, and Honey}]{tomita2021similarity}
Tomita, T.M., Barense, M.D., Honey, C.J.: The similarity structure of
  real-world memories. bioRxiv  (2021), \doi{10.1101/2021.01.28.428278},
  \urlprefix\url{https://www.biorxiv.org/content/early/2021/01/30/2021.01.28.428278}

\bibitem[{Waller and Anderson(2021)}]{waller2021quantifying}
Waller, I., Anderson, A.: Quantifying social organization and political
  polarization in online platforms. Nature \textbf{600}(7888), 264--268 (2021)

\bibitem[{Wang et~al.(2015)Wang, Hong, and Pi}]{wang2015crosscultural}
Wang, J., Hong, J.Z., Pi, Z.L.: Cross-{Cultural} {Adaptation}: {The} {Impact}
  of {Online} {Social} {Support} and the {Role} of {Gender}. Social Behavior
  and Personality: an international journal \textbf{43}(1), 111--121 (Feb
  2015), ISSN 0301-2212, \doi{10.2224/sbp.2015.43.1.111},
  \urlprefix\url{https://www.ingentaconnect.com/content/10.2224/sbp.2015.43.1.111}

\bibitem[{Wang et~al.(2013)Wang, Cai, Li, Jiang, Wang, Song, and
  Xia}]{wang2013optimal}
Wang, Y., Cai, H., Li, C., Jiang, Z., Wang, L., Song, J., Xia, J.: Optimal
  caliper width for propensity score matching of three treatment groups: A
  monte carlo study. PLOS ONE \textbf{8}(12), 1--7 (12 2013),
  \doi{10.1371/journal.pone.0081045},
  \urlprefix\url{https://doi.org/10.1371/journal.pone.0081045}

\bibitem[{Weld et~al.(2022)Weld, West, Glenski, Arbour, Rossi, and
  Althoff}]{weld2022adjusting}
Weld, G., West, P., Glenski, M., Arbour, D., Rossi, R.A., Althoff, T.:
  Adjusting for confounders with text: Challenges and an empirical evaluation
  framework for causal inference. Proceedings of the International AAAI
  Conference on Web and Social Media \textbf{16}(1), 1109--1120 (May 2022),
  \doi{10.1609/icwsm.v16i1.19362},
  \urlprefix\url{https://ojs.aaai.org/index.php/ICWSM/article/view/19362}

\bibitem[{Xi and Singh(2024)}]{xi2023blame}
Xi, R., Singh, M.P.: The blame game: Understanding blame assignment in social
  media. IEEE Transactions on Computational Social Systems \textbf{11}(2),
  2267--2276 (2024), \doi{10.1109/TCSS.2023.3261242}

\bibitem[{Yang et~al.(2018)Yang, Zhong, Kumar, Chow, and
  Ouyang}]{yang2018exchanging}
Yang, F., Zhong, B., Kumar, A., Chow, S.M., Ouyang, A.: Exchanging {Social}
  {Support} {Online}: {A} {Longitudinal} {Social} {Network} {Analysis} of
  {Irritable} {Bowel} {Syndrome} {Patients}' {Interactions} on a {Health}
  {Forum}. Journalism \& Mass Communication Quarterly \textbf{95}(4),
  1033--1057 (Dec 2018), ISSN 1077-6990, 2161-430X,
  \doi{10.1177/1077699017729815},
  \urlprefix\url{http://journals.sagepub.com/doi/10.1177/1077699017729815}

\bibitem[{Zhou et~al.(2017)Zhou, Heather, Cesare, and Ryder}]{zhou2017ask}
Zhou, B., Heather, D., Cesare, A.D., Ryder, A.G.: Ask and you might receive:
  {The} actor--partner interdependence model approach to estimating cultural
  and gender variations in social support. European Journal of Social
  Psychology \textbf{47}(4), 412--428 (Jun 2017), ISSN 0046-2772, 1099-0992,
  \doi{10.1002/ejsp.2251},
  \urlprefix\url{https://onlinelibrary.wiley.com/doi/10.1002/ejsp.2251}

\end{thebibliography}
\end{document}